\documentclass[aps,prd,onecolumn,showpacs,groupedaddress]{revtex4-2}
\usepackage{graphicx}
\usepackage{dcolumn}
\usepackage{bm}
\usepackage{color}
\usepackage{longtable}
\usepackage{supertabular}
\usepackage[T1]{fontenc}
\usepackage{epstopdf}
\usepackage{amsmath}
\usepackage{amssymb}
\usepackage{setspace}
\usepackage{multirow}
\usepackage{booktabs}
\usepackage[section]{placeins}
\usepackage{mathrsfs}
\usepackage{hyperref}

\makeatletter

\newcommand{\Rmnum}[1]{\expandafter\@slowromancap\romannumeral #1@} 
\newcommand{\bq}{\begin{equation}}
\newcommand{\eq}{\end{equation}}
\newcommand{\bqn}{\begin{eqnarray}}
\newcommand{\eqn}{\end{eqnarray}}
\newcommand{\nb}{\nonumber}

\makeatother

\begin{document}

\title{An MLE analysis on the relationship between the initial-state granularity and final-state flow factorization}

\author{Shui-Fa Shen$^{1,2}$}
\author{Chong Ye$^{3}$}
\author{Dan Wen$^{4}$}
\author{Lina Bao$^{5}$}
\author{Jin Li$^{6}$}
\author{Yutao Xing$^{7}$}
\author{Jiaming Jiang$^{8}$}\email[E-mail: ]{jiangjiaming@ecit.cn (corresponding author)}
\author{Wei-Liang Qian$^{9,3}$}\email[E-mail: ]{wlqian@usp.br (corresponding author)}

\affiliation{$^{1}$ School of Intelligent Manufacturing, Zhejiang Guangsha Vocational and Technical University of Construction, 322100, Jinhua, Zhejiang, China}
\affiliation{$^{2}$ Hefei Institutes of Physical Science, Chinese Academy of Sciences, 230031, Hefei, Anhui, China}
\affiliation{$^{3}$ Center for Gravitation and Cosmology, College of Physical Science and Technology, Yangzhou University, 225009, Yangzhou, China}
\affiliation{$^{4}$ College of Science, Chongqing University of Posts and Telecommunications, 400065, Chongqing, China}
\affiliation{$^{5}$ Department of Basic Sciences, Army Academy of Artillery and Air Defense, 230031, Hefei, Anhui, China}
\affiliation{$^{6}$ College of Physics, Chongqing University, 401331, Chongqing, China}
\affiliation{$^{7}$ Instituto de F\'isica, Universidade Federal Fluminense, 24210-346, Niter\'oi, RJ, Brazil}
\affiliation{$^{8}$ College of Nuclear Science and Engineering, East China University of Technology, 330013, Nanchang, Jiangxi, China}
\affiliation{$^{9}$ Escola de Engenharia de Lorena, Universidade de S\~ao Paulo, 12602-810, Lorena, SP, Brazil}

\begin{abstract}
In this study, we employ the maximum likelihood estimator (MLE) to investigate the relationship between initial-state fluctuations and final-state anisotropies in relativistic heavy-ion collisions.
The granularity of the initial state, reflecting fluctuations in the initial conditions (IC), is modeled using a peripheral tube model.
Besides differential flow, our analysis focuses on a class of more sensitive observables known as flow factorization.
Specifically, we evaluate these observables using MLE, an asymptotically normal and unbiased tool in standard statistical inference.
Our findings show that the resulting differential flow remains essentially unchanged for different IC defined by the peripheral tube model.
The resulting harmonic coefficients obtained using MLE and multi-particle cumulants are found to be consistent.
However, the calculated flow factorizations show significant variations depending on both the IC and the estimators, which is attributed to their sensitivity to initial-state fluctuations.
Thus, we argue that MLE offers a compelling alternative to standard methods such as multi-particle correlators, particularly for sensitive observables constructed from higher moments of the azimuthal distribution.
\end{abstract}

\date{Nov. 20th, 2024}

\maketitle
\newpage

\section{Introduction}\label{section1}

Relativistic hydrodynamics is a widely accepted theoretical framework for modeling the temporal evolution of strongly coupled quark-gluon plasmas produced in relativistic heavy-ion collisions~\cite{hydro-review-04, hydro-review-05, hydro-review-06, hydro-review-07, hydro-review-08, hydro-review-09, hydro-review-10}.
This macroscopic approach treats the plasma as a continuum, which is crucial for capturing the underlying physics that gives rise to observable phenomena.
Key observables in relativistic hydrodynamics include the particle spectrum at intermediate and low transverse momentum, with particular emphasis on collective properties such as flow harmonics and correlations~\cite{Ollitrault:1992bk, Voloshin:1994mz, Ollitrault:1997vz, Borghini:2000sa, Takahashi:2009na, Andrade:2010xy, Luzum:2010sp}.
Experimentally, measurements of azimuthal distributions have been instrumental in demonstrating the concept of a {\it perfect} liquid, first observed at RHIC~\cite{STAR:2000ekf}.
Consequently, azimuthal anisotropy has emerged as a crucial observable for extracting information about the properties of the underlying physical system~\cite{BRAHMS:2004adc, PHOBOS:2004zne, STAR:2005gfr, ATLAS:2012at, CMS:2012zex, CMS:2012tqw}.
Recently, these observables have been further studied in the context of deformed nuclei~\cite{Holtermann:2023vwr, ATLAS:2014qxy, ATLAS:2019peb}.

The essence of hydrodynamical evolution can be viewed mainly as the dynamic response to fluctuating initial conditions (IC). 
Given the inherently nonlinear nature of hydrodynamics, numerous studies have investigated these complex interactions.  
In particular, significant research has focused on understanding the relationship between initial-state eccentricities and the final-state azimuthal anisotropies~\cite{hydro-v3-02,hydro-vn-33,sph-vn-03,hydro-vn-34,sph-vn-04,sph-vn-06,hydro-vn-45,sph-corr-30}.  
The quantitative decomposition of anisotropic IC was first introduced in Refs.~\cite{hydro-v3-02,hydro-vn-33}.  
This approach relies on a cumulant expansion, where each expansion coefficient captures the ``connected'' eccentricity component at a given order.  
Consequently, higher-order cumulants are defined by subtracting contributions from the ``disconnected'' combinations of lower-order terms.  
Furthermore, flow harmonics represent the hydrodynamic response to IC fluctuations, classified according to their cumulant order, with the lowest-order cumulants generally assumed to have the most significant impact.  

In the literature, contributions proportional to cumulants of the same azimuthal order are often identified as linear responses.  
In contrast, contributions from combinations of lower-order cumulants that result in the same azimuthal order are attributed to non-linearity.  
In practice, the response strength varies across different cumulant combinations.  
Specifically, the established mapping between IC and flow harmonics is defined by the strongest correlation between an optimized linear combination of cumulant products and the corresponding flow harmonic~\cite{sph-vn-03,sph-vn-06}.  
Numerical studies confirm that this mapping is particularly effective for more central collisions, inspiring further research in this area~\cite{hydro-vn-34,hydro-vn-45,sph-corr-30}.  
For example, one might examine whether each individual component of a given azimuthal order results in a linear response by isolating them from the IC~\cite{sph-vn-04}.
It should be noted that each azimuthal harmonic order corresponds to an infinite set of moments or cumulants.  
Therefore, it remains unclear whether a specific one-to-one mapping exists or to what extent different terms of the same azimuthal order mix during dynamical evolution.  
Consequently, this ambiguity leaves room for alternative descriptions of the radial expansion.  

In this regard, some authors argued that the cumulant expansion may not be optimal for decomposing the IC.
Alternatively, the Bessel-Fourier expansion proposed in Refs.~\cite{hydro-vn-ph-09,hydro-vn-ph-10} utilizes an orthonormal basis to decompose IC fluctuations.
A vital advantage of this approach is that it orders fluctuations based on their wavelength, allowing shorter wavelength radial modes to be effectively suppressed.
This enables a more precise separation of different modes within a hydrodynamic framework~\cite{hydro-vn-ph-11}.
Another intuitive method for capturing the granularity of the IC is the {\it peripheral tube model}~\cite{sph-corr-02, sph-corr-04, sph-corr-18, sph-review-02}, which provides an intuitive interpretation for the generation of triangular flow and ridge structures observed in di-hadron correlations.
Unlike Fourier-based eccentricity decompositions, this model aims to capture the localized feature in the event-by-event fluctuating IC.
It is motivated by heuristic arguments and numerical simulations, which suggest that high-energy-density peaks are formed near the surface region due to elementary binary collisions in the transverse plane.
These localized regions naturally form a tube-like structure along the longitudinal direction and are shown to correlate strongly with the observed ridge-like structures in two-particle correlations.
This model emphasizes the impact of localized IC fluctuations, as opposed to global sinusoidal expansions based on the azimuthal moment expansion.
For instance, a tube located deep inside the medium, though it has a sizable contribution to the corresponding moments, would have its hydrodynamic effect significantly suppressed by the surrounding matter.
In contrast, a tube near the surface can induce notable distortions in the single-particle azimuthal distribution and influence two-particle correlations.
This model has been successfully employed to study various features in di-hadron correlations, showing good agreement with experimental data~\cite{sph-corr-02, sph-corr-12, sph-corr-13, sph-review-02}.

The anisotropic distribution of final-state particles in momentum space is characterized by flow harmonics $v_n$, defined through the one-particle distribution function~\cite{Voloshin:1994mz, Voloshin:2008dg}: \begin{eqnarray}
f_1(\phi)=\frac{1}{2\pi}\left[1+\sum_{n=1}^{} 2v_{n}\cos{n(\phi-\Psi_n)}\right],
\label{oneParDis}
\end{eqnarray}
where $\phi$ is the azimuthal angle of the emitted particle and $\Psi_n$ represents the event plane for the harmonic order $n$.
Elliptic flow ($v_2$) and triangular flow ($v_3$) are particularly relevant observables, with $v_2$ primarily arising from the almond-shaped geometry of the overlap region~\cite{Ollitrault:1992bk} and $v_3$ stemming from event-by-event IC fluctuations~\cite{Alver:2010gr}.
Considerable research has focused on exploring the relationship between initial geometric anisotropies and final-state flow harmonics, primarily to investigate non-linear effects~\cite{Teaney:2012ke, Niemi:2012aj, Qian:2013nba, Yan:2014nsa, Fu:2015wba, Yan:2015jma, Wen:2020amn}, eccentricity and flow fluctuations~\cite{Hama:2007dq, Bhalerao:2011yg, Heinz:2013bua, Gronqvist:2016hym}, and multi-particle correlations~\cite{Bhalerao:2013ina, Denicol:2014ywa}.

Various techniques have been developed to determine flow harmonics $v_n$ from experimental data.
The traditional event plane method~\cite{Voloshin:1994mz, Poskanzer:1998yz} estimates the event plane angles $\Psi_n$ to evaluate the harmonics in Eq.~\eqref{oneParDis}, reflecting the fact that the reaction plane\cite{Alver:2010gr} cannot be directly measured.
Other methods, such as particle correlations, use Q-vectors and cumulants~\cite{Danielewicz:1985hn, Borghini:2000sa, Bilandzic:2010jr, Jia:2017hbm} to eliminate the need for $\Psi_n$.
This approach allows cumulants to be compactly expressed using generating functions~\cite{Borghini:2000sa, Borghini:2001vi}.
It includes techniques like particle cumulants~\cite{Borghini:2000sa, Borghini:2001vi, Bilandzic:2010jr}, Lee-Yang zeros~\cite{Bhalerao:2003xf, Bhalerao:2003yq}, and symmetric cumulants~\cite{Bilandzic:2013kga}.
Recently, an alternative method based on the maximum likelihood estimator (MLE) was proposed~\cite{sph-vn-10}, treating the flow coefficients as unknown parameters of a hypothetical probability distribution derived from experimental data.
The MLE approach offers several advantages.
First, it effectively handles nonflow effects, which refer to correlations that cannot be attributed to collective behavior as per Eq.\eqref{oneParDis}.
Momentum conservation, for example, can cause deviations from a purely flow-driven spectrum~\cite{Chajecki:2008yi}.
Second, MLE is efficient at the limit of large samples and can incorporate additional constraints in the flow distribution function, making it robust in scenarios with nonflow effects.
Furthermore, MLE's asymptotic normality ensures unbiased or nearly unbiased results, making it well-suited for the extensive datasets obtained at RHIC and LHC.
Given these attributes, MLE provides a unique tool for analyzing multi-particle correlations and assessing the connection between initial anisotropy and final-state flow harmonics.

Beyond flow harmonics, the discussion can be extended to more complex observables constructed using generic multi-particle correlators.
As highlighted in the literature~\cite{ALICE:2011svq, Gardim:2012im}, such quantities are often more sensitive to the underlying IC's specific characteristics than their global averaged properties.
Specifically, event-by-event fluctuations in the IC can introduce significant effects in multi-particle correlators, leading to transverse momentum ($p_{\rm T}$) dependent event planes~\cite{Heinz:2013bua}.
To capture these variations, the final-state event planes are typically estimated using the azimuthal distribution of particles over a broad $p_{\rm T}$ range on an event-by-event basis.
Experimental data~\cite{ALICE:2022dtx, Zhou:2014bba} and hydrodynamic simulations~\cite{Alver:2010gr, Heinz:2013bua} indicate that the event planes fluctuate significantly across different $p_{\rm T}$ intervals.
This observation, which is supported by studies such as Refs.~\cite{CMS:2015xmx, ALICE:2017lyf, Barbosa:2021ccw}, reveals that the correlation matrix in transverse momentum approximately factorizes.
Such factorization is considered compelling evidence for the hydrodynamic picture of heavy-ion collisions.
Consequently, flow factorization~\cite{ALICE:2011svq, Gardim:2012im, Heinz:2013bua} has emerged as a powerful tool for probing the properties of initial-state fluctuations.

The breakdown of flow factorization is primarily attributed to event-by-event fluctuations in the initial energy distribution~\cite{ALICE:2011svq, Gardim:2012im, Heinz:2013bua, CMS:2015xmx, ALICE:2017lyf} rather than to variations in the transport properties of the medium.
This suggests that a detailed analysis of factorization breakdown can yield valuable insights into the IC of the system.
Moreover, distinct differences have been observed between flow Pearson correlations constructed using different moments~\cite{sph-vn-11}.
Given the ability of the MLE to preserve the structure of these correlations owing to its equivariance properties, it provides a robust and meaningful framework for assessing multi-particle correlators and their sensitivity to initial-state granularity.

The present study is motivated by the above considerations.
We aim to explore the relationship between initial-state anisotropies and final-state flow harmonics and their correlations by employing MLE as a statistical estimator for the flows and other higher-order moments.
Specifically, we use the peripheral tube model to quantify IC granularity and employ flow factorization to capture nuances in high-order flow harmonics and multi-particle correlators.
In particular, MLE has been applied to specific correlators that are otherwise inaccessible for an arbitrary combination.

The remainder of this paper is organized as follows.
In the next section, we discuss MLE as a statistical estimator for multi-particle correlators and flow factorization.
In Sec.~\ref{section3}, we introduce the peripheral tube model and describe its application in modeling ICs with varying granularity.
Numerical studies are carried out in Sec.~\ref{section4}, where we investigate the relationship between IC granularity and final-state flow factorization.
The results are compared with those obtained using cumulants and the event-plane method.
The final section provides further discussions and concluding remarks.

\section{Statistical estimators for multi-particle correlator and flow factorization}\label{section2}

The most prominent methods for extracting flow harmonics are based on multi-particle correlations~\cite{Poskanzer:1998yz}.
These techniques rely on the following definition for $k$-particle correlator~\cite{Bhalerao:2011yg}: 
\begin{eqnarray}
\langle k\rangle_{n_1,\cdots,n_k} \equiv \langle e^{i(n_1\phi_1 + \cdots + n_k\phi_k)} \rangle
= v_{n_1}\cdots v_{n_k} e^{i\left(n_1\Psi_{n_1}+\cdots+n_k\Psi_{n_k}\right)} , \label{nkCorr}
\end{eqnarray}
where $\langle\cdots\rangle$ denotes the average over distinct tuples of particles, assuming independent particle emission as described by Eq.~\eqref{oneParDis} in the limit of infinite multiplicity.

To isolate $v_n$, one typically chooses a specific set of indices $(n_1, \cdots, n_k)$ such that~\cite{Bhalerao:2011yg}
\begin{eqnarray}
\sum_{j=1}^k n_j = 0 ,\label{sumRes}
\end{eqnarray}
which ensures that all contributions from the event planes cancel in the exponential.
This condition effectively reduces the expression to a formalism independent of the event-plane angles~\cite{Bhalerao:2013ina}.
The simplest example is the two-particle correlation ($k=2$), where $n_1 = -n_2 = n$.
This choice yields
\begin{eqnarray}
\langle 2 \rangle_{n,-n} \equiv \left \langle e^{in(\phi_{1}-\phi_{2})} \right \rangle
= \langle\cos{n(\phi_{1}-\phi_{2})} \rangle = v_n^2 ,
\label{eq2}
\end{eqnarray}
which directly relates the two-particle correlation to the square of the flow harmonic $v_n$.

In realistic scenarios, however, the number of particles ($M$) in a given event is finite.
Thus, the analysis is performed using discrete azimuthal angles $\phi_1, \phi_2, \cdots, \phi_M$ corresponding to the measured particles.
Rather than employing the integration in Eq.~\eqref{eq2}, which is not viable for a finite number of particles $M$, it is intuitive to use the following summation~\cite{Bilandzic:2013kga}: 
\begin{eqnarray}
\widehat{v_n^2} = \frac{1}{M(M-1)}\sum_{i\ne j} \cos n(\phi_i - \phi_j) ,
\label{eqEst2}
\end{eqnarray}
to estimate the flow harmonic $v_n$.

As a more sensitive observable, flow factorization is defined as a Pearson correlation in terms of flow vectors evaluated at different transverse momenta, namely, 
\begin{eqnarray}
       r_n(p_{\rm T}^{\rm a},p_{\rm T}^{\rm t})=\frac{V_{n\Delta}(p_{\rm T}^{\rm a},p_{\rm T}^{\rm t})}{\sqrt{V_{n\Delta}(p_{\rm T}^{\rm a},p_{\rm T}^{\rm a})V_{n\Delta}(p_{\rm T}^{\rm t},p_{\rm T}^{\rm t})}} ,
\label{rn_fact}
\end{eqnarray}
where $p_{\rm T}^{\rm t}$ and $p_{\rm T}^{\rm a}$ are transverse momenta of the trigger and associated particles, and $V_{n\Delta}(p_{\rm T1},p_{\rm T2})$ is the $n$th harmonic of the underlying di-hadron azimuthal distribution with transverse momenta $p_{\rm T1}$ and $p_{\rm T2}$, namely,
\begin{eqnarray}
       V_{n\Delta}(p_{\rm T1},p_{\rm T2})
       \equiv \left\langle e^{in\left(\phi(p_{\rm T1})-\phi(p_{\rm T2})\right)}\right\rangle 
       =\langle \cos n\left(\phi(p_{\rm T1}) -\phi(p_{\rm T2})\right) \rangle 
       =\langle V_n^*(p_{\rm T1})V_n(p_{\rm T2})\rangle ,
       \label{DefVnDelta}
\end{eqnarray}
where 
\bqn
V_n(p_{\rm T})=v_n(p_{\rm T}) e^{-in\Psi_n(p_{\rm T})},\label{VnDeltaFactorization}
\eqn 
is known as the flow vector~\cite{Luzum:2013yya, Ollitrault:2012cm}. 
Because of its explicit consideration of transverse momentum dependence, Eq.~\eqref{DefVnDelta} can be viewed as the {\it differential} counterpart of the two-particle correlation defined in Eq.~\eqref{eq2}.
Similar to Eq~\eqref{eqEst2}, in practice, Eq.~\eqref{DefVnDelta} is implemented by enumerating all possible combinations of the measured particles, as given in the Appx.~A of~\cite{sph-vn-11}

From the statistical inference perspective, Eq.~\eqref{eqEst2} serves as an estimator, providing an estimation for $v_n^2$ rather than $v_n$ itself, based on a finite number of observations.
The first two moments of this estimator can be readily evaluated~\cite{sph-vn-10} and generally do not vanish.
Specifically, this estimator has a finite variance that decreases with increasing multiplicity.
For higher-order correlators, one can employ generating functions~\cite{Borghini:2000sa, Borghini:2003ur, Borghini:2007ku} or compute them directly~\cite{Bilandzic:2010jr} using $Q$-vectors~\cite{Danielewicz:1985hn} or flow vectors~\cite{Luzum:2013yya, Ollitrault:2012cm}.
Notably, quantities constructed using $Q$-vectors or flow vectors can also be interpreted as estimators, serving as generalized extensions of the unweighted sums in Eq.~\eqref{nkCorr}.
Notably, the variance of these quantities largely remains finite~\cite{sph-corr-31}, indicating an inherent statistical uncertainty due to the finite number of measured events and particle multiplicity.
A finite variance implies that uncertainties in the estimated flow harmonics are inevitably influenced by limited statistics, particularly for a finite number of events.
Therefore, such limitations of statistical origin must be considered when comparing results obtained from different flow measurement methods.

Following this line of thought, MLE could serve as an alternative estimator for flow and related observables, a possibility explored in a previous study~\cite{sph-vn-10}.
Expressly, for a given set of observations $y\equiv (y_1, y_2, \cdots, y_M)$, we assume that they are sampled from a joint probability distribution governed by several unknown parameters $\theta \equiv (\theta_1, \theta_2, \cdots, \theta_m)$.
As mentioned in the Introduction, the likelihood function $\mathcal{L}$ for the observed data is given by:
\begin{eqnarray}
\mathcal{L}(\theta) \equiv \mathcal{L}(\theta; y) = f(y; \theta).\label{defLn}
\end{eqnarray}
which represents the joint probability density for the given observations evaluated at the parameters $\theta$.
The goal of MLE is to determine the parameters for which the observed data attains the highest joint probability, namely:
\begin{eqnarray}
\hat{\theta}_{\mathrm{MLE}}=\arg\max\limits_{\theta\in\Theta}\mathcal{L}(\theta) , \label{defMLE}
\end{eqnarray}
where $\Theta$ is the domain of the parameters.
In particular, for independent and identically distributed (i.i.d.) random variables, $f(y; \theta)$ is given by a product of likelihood functions $f^\mathrm{uni}$:
\begin{eqnarray}
f(y; \theta)=\prod_{j=1}^M f^\mathrm{uni}(y_j; \theta) . \label{iidfn1}
\end{eqnarray}

This scheme can be readily applied to collective flow in heavy-ion collisions.
Considering an event consisting of $M$ particles, the likelihood function reads:
\begin{eqnarray}
\mathcal{L}(\theta; \phi_{1}, \cdots, \phi_M) = f(\phi_{1},\cdots, \phi_M; \theta)=\prod_{j=1}^{M}f_1(\phi_j; \theta) ,
\label{eqlikelihood}
\end{eqnarray}
where the likelihood function $\mathcal{L}$ is governed by the one-particle distribution function Eq.~\eqref{oneParDis}.
The last equality is based on the assumption that the particles' azimuthal angles are i.i.d. variables.
The parameters of the distribution, $\theta = (v_1, v_2, \cdots, \Psi_1, \Psi_2, \cdots)$, are the flow harmonics and the event planes.

In practice, one often chooses the objective function to be the log-likelihood function ${\ell}$:
\begin{eqnarray}
{\ell}(\theta; \phi_{1}, \cdots, \phi_M) = \log\mathcal{L}(\theta; \phi_{1}, \cdots, \phi_M)
= \sum_{j=1}^{M}\log f_1(\phi_j; \theta) .
\label{eqlogl}
\end{eqnarray}
Numerical calculations indicate that Eq.~\eqref{eqlogl} is more favorable than Eq.~\eqref{eqlikelihood}, although as multiplicity $M$ increases, an appropriate implementation should be adopted to avoid the increasing truncation error.

The maximum of $\ell$ occurs at the same value of $\theta$, which maximizes $\mathcal{L}$.
For $\ell$ that is differentiable in its domain $\Theta$, the necessary conditions for the occurrence of a maximum are:
\begin{eqnarray}
\frac{\partial{\ell}}{\partial\theta_1}=\cdots=\frac{\partial{\ell}}{\partial\theta_m}=0 .
\label{condMLE}
\end{eqnarray}

As discussed in the Introduction, MLE has asymptotic normality, which attains the Cramér-Rao lower bound as the sample size increases.
In other words, no consistent estimator has a lower asymptotic mean squared error than the MLE.
In the context of relativistic heavy-ion collisions, all events of a given multiplicity asymptotically form a (multivariate) normal distribution:
\begin{eqnarray}
\hat{\theta}_{\mathrm{MLE}} \sim N\left(\theta_0, (I_M(\theta_0))^{-1}\right) ,
\label{assNormMLE}
\end{eqnarray}
where $\theta_0$ represents the true value, and $I_M(\theta)$ is the Fisher information matrix, defined as:
\begin{eqnarray}
I_M(\theta) \equiv E_\theta\left[-\frac{d^2}{d\theta^2}{\ell}(\theta;\phi_1,\cdots,\phi_M)\right] ,
\label{defIM}
\end{eqnarray}
where the expectation $E_\theta$ is taken with respect to the distribution $f(\phi_1,\cdots,\phi_M;\theta)$.
For i.i.d. data, the Fisher information possesses the form:
\begin{eqnarray}
I_M(\theta) = M I_1(\theta),
\label{defI1}
\end{eqnarray}
where $I_1$ is the Fisher information matrix for a single observation.
As a result, the standard deviation of MLE is expected to be roughly proportional to $\frac{1}{\sqrt{M}}$.
As discussed in~\cite{sph-vn-10}, these properties can be further quantified using the Wald, likelihood ratio, and score tests.

\section{The peripheral tube model}\label{section3}

The peripheral tube model~\cite{sph-review-02} offers a simplified yet effective framework for understanding the generation of triangular flow and the resulting particle correlations.
This approach is intrinsically connected to event-by-event fluctuating hydrodynamics, where the IC are represented by a few high-energy tubes near the system's surface.
Each of these high-energy tubes independently influences the local hydrodynamic evolution, and these tubes' collective contributions combine to produce the observed particle correlations.

The present study makes use of the tube model to quantify the granularity of the fluctuating IC, while the bulk of the hot matter is substituted by an average energy density distribution obtained from multiple events of the same centrality class.
We systematically vary the size and number of these tubes to investigate their impact on flow harmonics and particle correlations, focusing on flow factorization.
The IC in this model is devised to reflect the underlying event-by-event fluctuating initial energy density distribution, consisting of two main components: a smooth background and several high-energy-density tubes close to the surface.
The smooth background captures the bulk properties of the system, while the tubes represent localized fluctuations on an event-by-event basis.
The energy density profile in the model is expressed as: 
\begin{eqnarray}
\epsilon = \epsilon_\mathrm{bgd}+\epsilon_\mathrm{tube} , 
\end{eqnarray}
where the average background distribution $\epsilon_\mathrm{bgd}$ is given by: 
\begin{eqnarray}
\epsilon_\mathrm{bgd} = (K + Lr^2 + Mr^4) e^{-r^{2c}} ,  \label{avaIC}
\end{eqnarray}
where the radial coordinate is defined as: 
\begin{eqnarray}
r = \sqrt{ax^2 + by^2} , 
\end{eqnarray}
where $K,~L,~M,~a,~b,c$ are parameters obtained by numerical calibration to the average ICs generated by the NeXuS and EPOS models~\cite{nexus-1,nexus-rept,epos-1,epos-2,epos-3}.

The profile of an individual high-energy tube is given by: 
\begin{eqnarray}
\epsilon_\mathrm{tube} &=& A_\mathrm{tube}\exp\left[-\frac{(x-x_\mathrm{tube})^2+(y-y_\mathrm{tube})^2}{R_\mathrm{tube}^2}\right] ,  \label{energytube}
\end{eqnarray}
where $A_\mathrm{tube}$ and $R_\mathrm{tube}$ denote the maximum energy density and radius of the tube, respectively.
The radial position $r_\mathrm{tube}$ is defined as: 
\begin{eqnarray}
r_\mathrm{tube} &=& \frac{r_0}{\sqrt{a\cos^2\theta + b\sin^2\theta}} , 
\end{eqnarray}
with the spatial coordinates $(x_\mathrm{tube}, y_\mathrm{tube})$ given by: 
\begin{eqnarray}
x_\mathrm{tube} &=& r_\mathrm{tube}\cos\theta ,  \nonumber \\
y_\mathrm{tube} &=& r_\mathrm{tube}\sin\theta ,  \nonumber
\end{eqnarray}
where $r_0$, $a$, $b$, and $\theta$ determine the radial location and orientation of the tube.
For the present study, the number of tubes will be varied to reflect the different degrees of granularity of the IC.
Previous analysis suggests that this parameter has a limited effect on the resulting flow harmonics and two-particle correlations~\cite{sph-corr-20}.
Nonetheless, as shown below, different granularity has a rather subtle effect on the resulting collect flows.
The parameters used in this model are calibrated using typical IC profiles for Au+Au collisions in the $0\%-5\%$ centrality class at $\sqrt{s_{NN}} = 200$ GeV.
For a randomly generated event, the radii and azimuthal angles of the tubes are drawn from the following uniform distributions $r_0 \sim \mathrm{U}(0, 0.546)$ and $\theta \sim \mathrm{U}(0, 2\pi)$.
The remaining parameters are summarized in Tab.~\ref{tubeparameters}.

\begin{table}[h!]
\begin{center}
  \caption{The model parameters of the peripheral tube model employed in the present study.}
  \begin{tabular}{cccccccc}  
  \hline\hline
    ~~~~$K$~~~~&~~~~$L$~~~~&~~~~$M$~~~~&~~~~$a$~~~~&~~~~$b$~~~~&~~~~$c$~~~~&~~~~$A_\mathrm{tube}$~~~~&~~~~$R_\mathrm{tube}$~~~~\\ 
   \hline
   9.33 & 7.0 & 2.0 & 0.41 & 0.186 & 0.9 & 12.0 & 2.3\\
  \hline \hline
  \end{tabular}
  \label{tubeparameters}
\end{center}
\end{table}

\begin{figure}[ht]
    \centering
    \begin{minipage}{0.32\textwidth}
        \centering
        \includegraphics[width=1.15\textwidth, height=0.23\textheight]{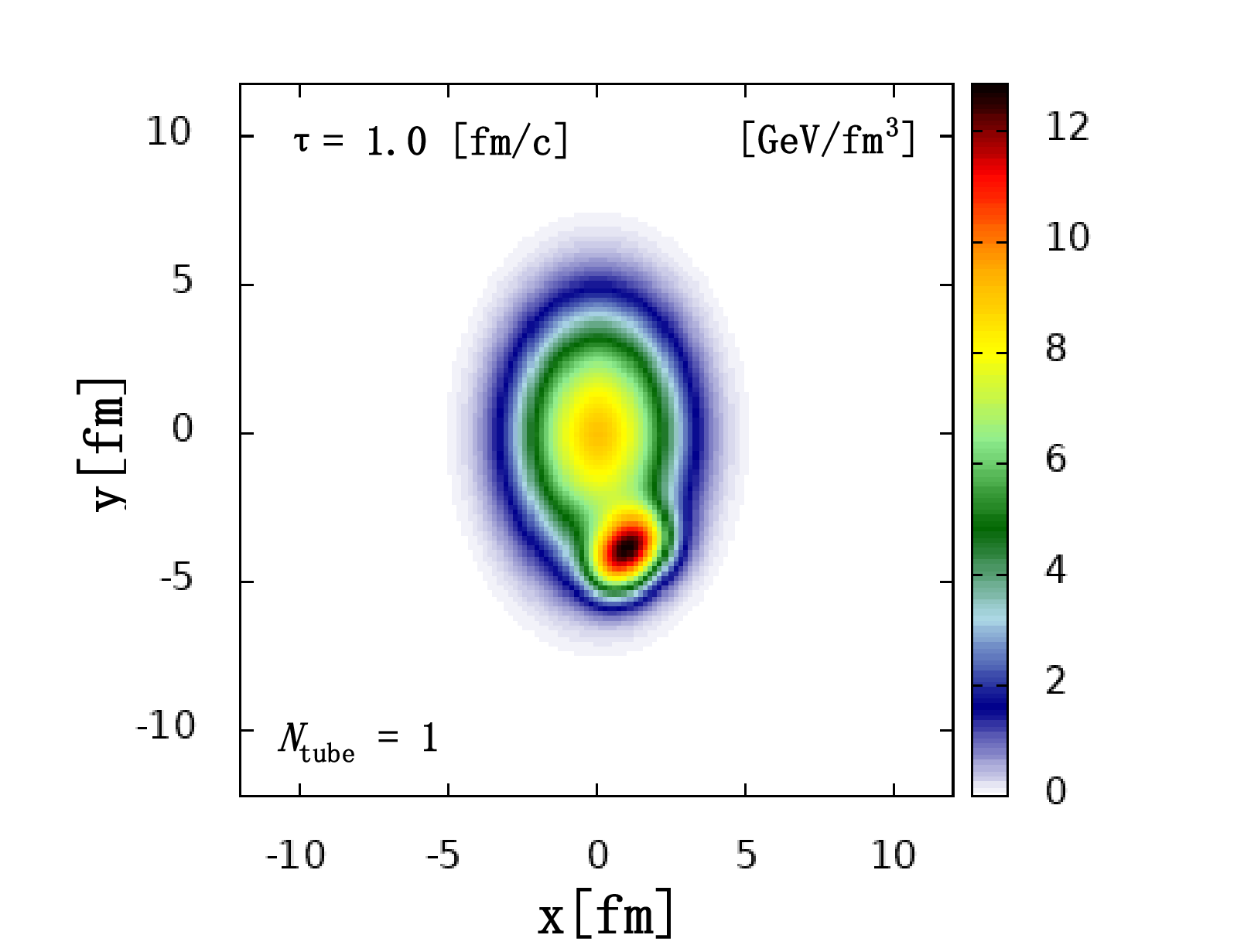}
    \end{minipage}
    \begin{minipage}{0.32\textwidth}
        \centering
        \includegraphics[width=1.15\textwidth, height=0.23\textheight]{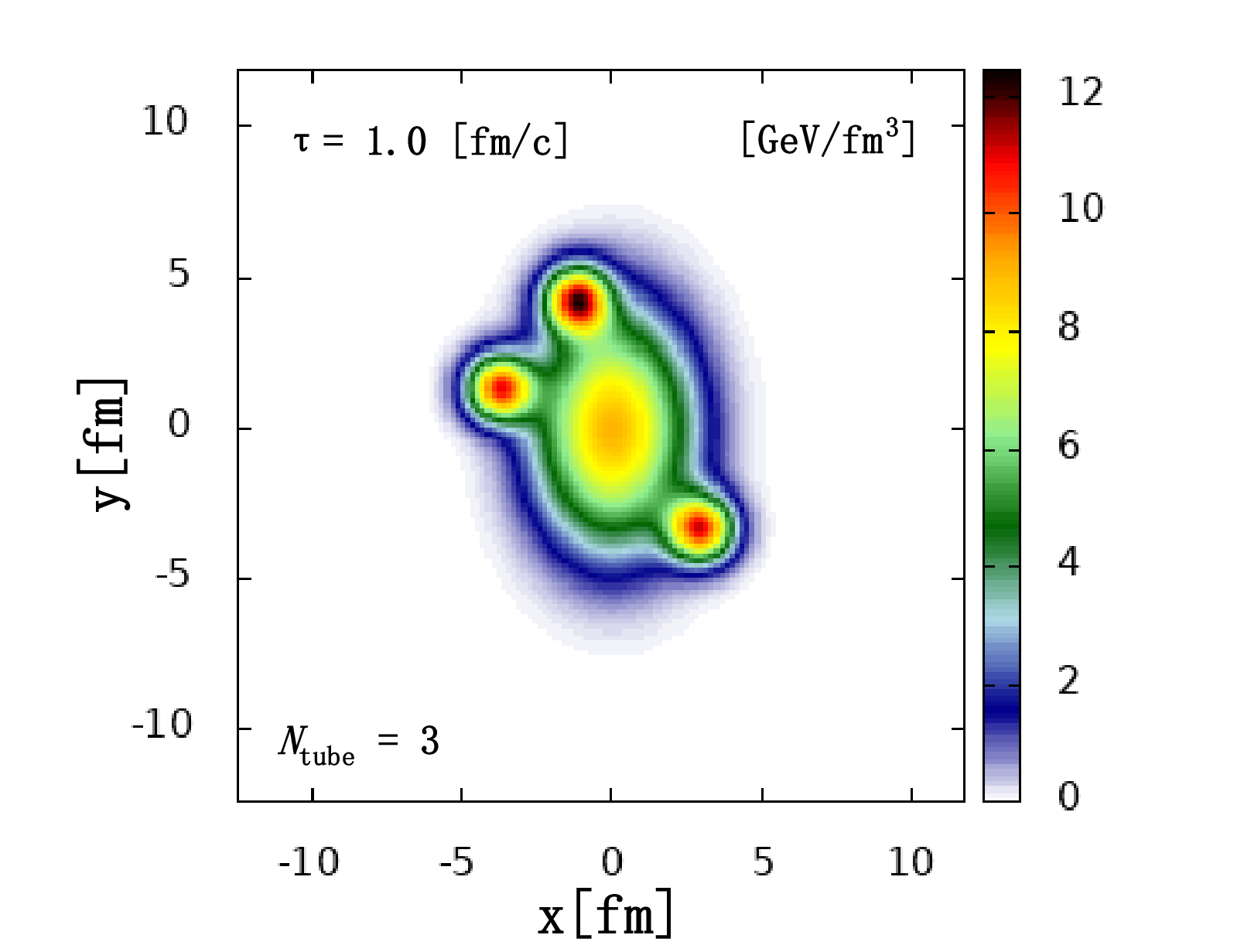}
    \end{minipage}
    \begin{minipage}{0.32\textwidth}
        \centering
        \includegraphics[width=1.15\textwidth, height=0.23\textheight]{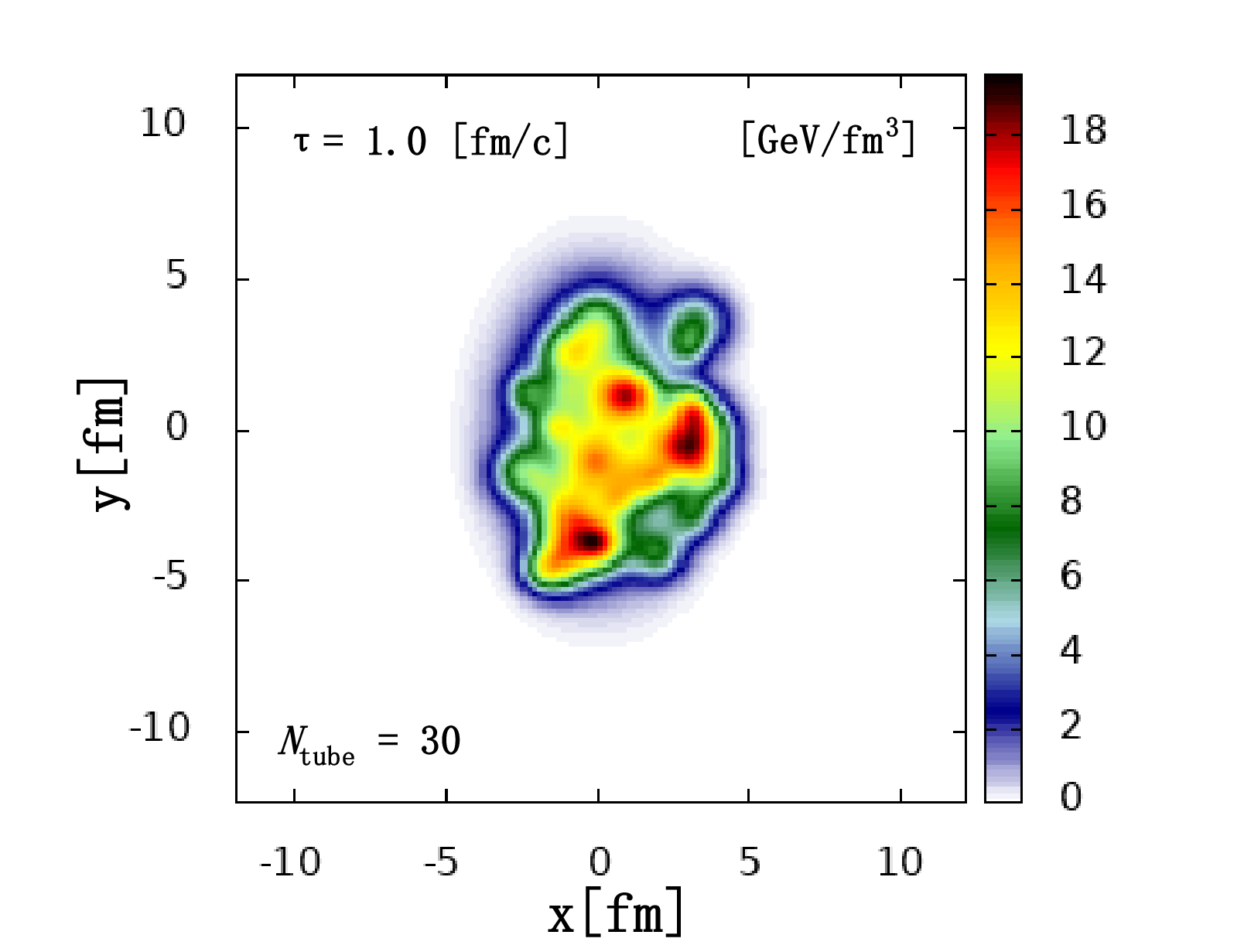}
    \end{minipage}
    \begin{minipage}{0.32\textwidth}
        \centering
        \includegraphics[width=1.15\textwidth, height=0.23\textheight]{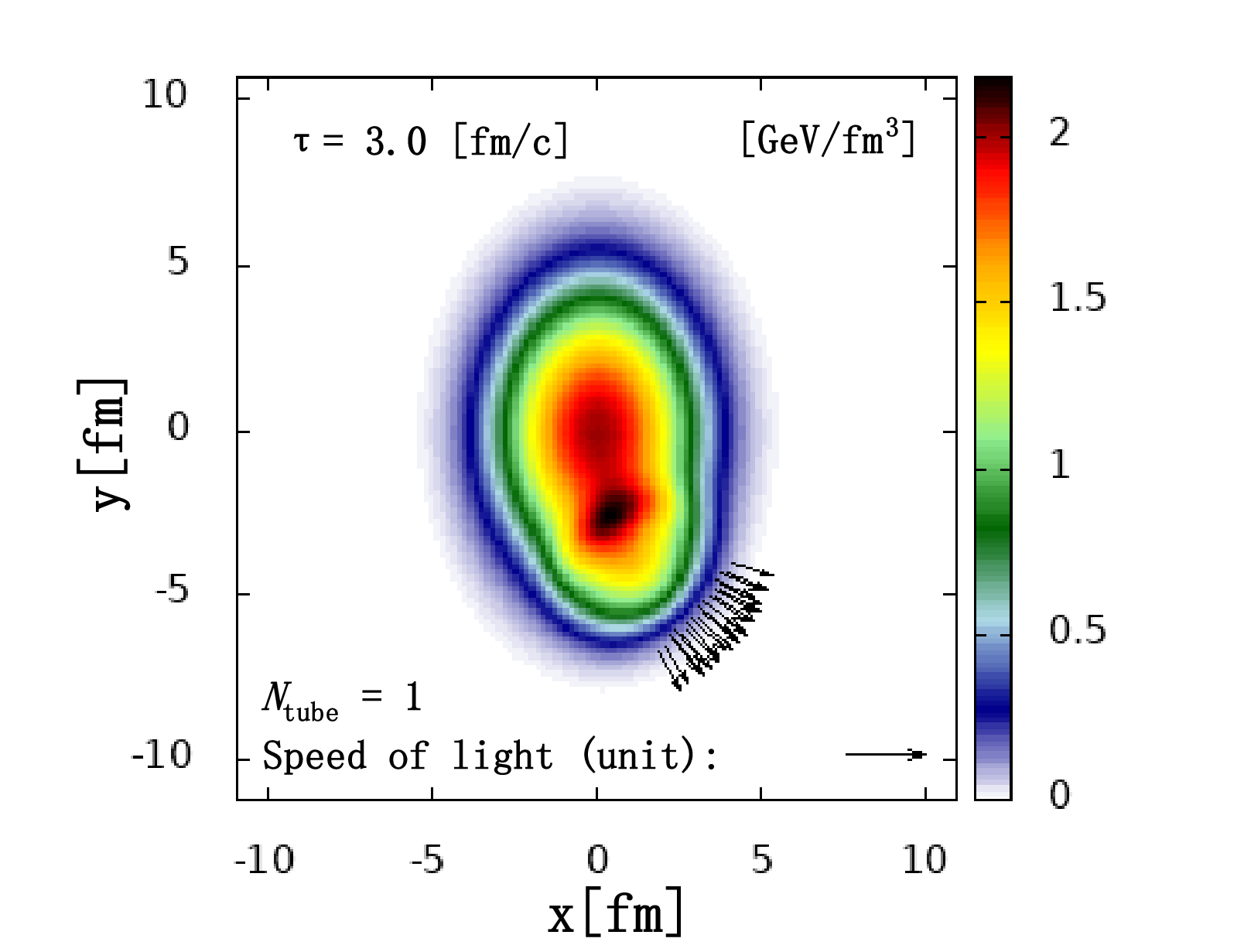}
    \end{minipage}
    \begin{minipage}{0.32\textwidth}
        \centering
        \includegraphics[width=1.15\textwidth, height=0.23\textheight]{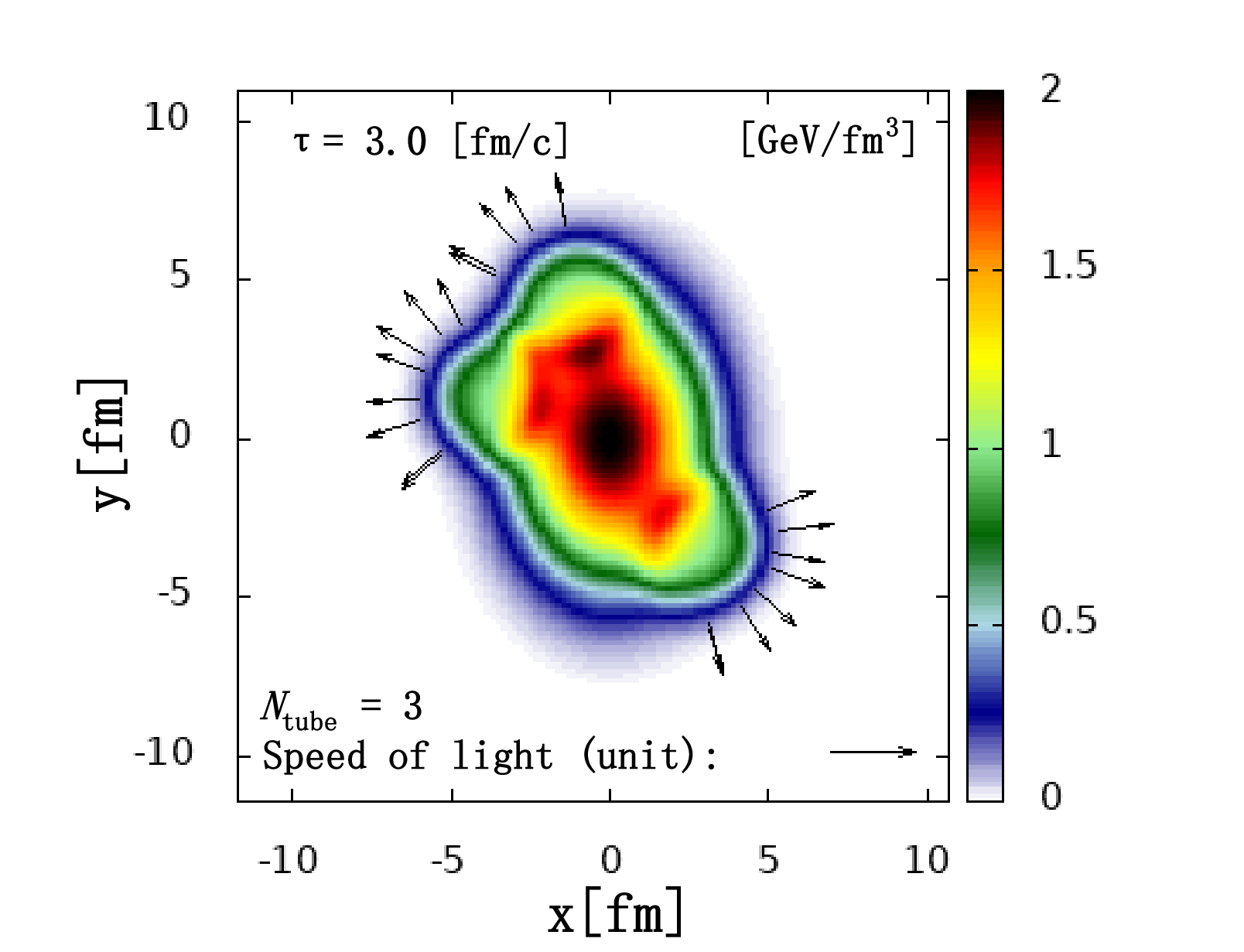}
    \end{minipage}
    \begin{minipage}{0.32\textwidth}
        \centering
        \includegraphics[width=1.15\textwidth, height=0.23\textheight]{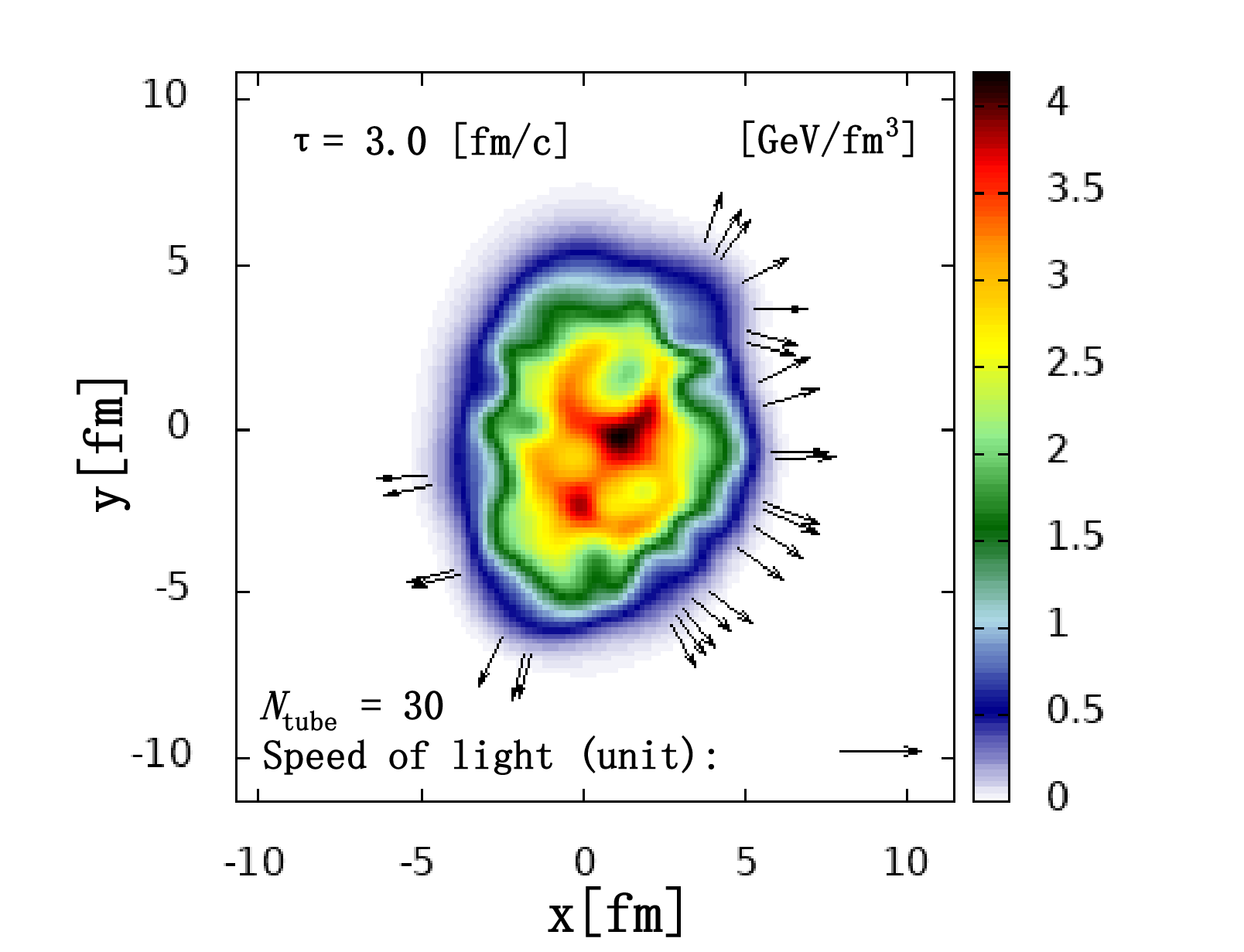}
    \end{minipage}
    \begin{minipage}{0.32\textwidth}
        \centering
        \includegraphics[width=1.15\textwidth, height=0.23\textheight]{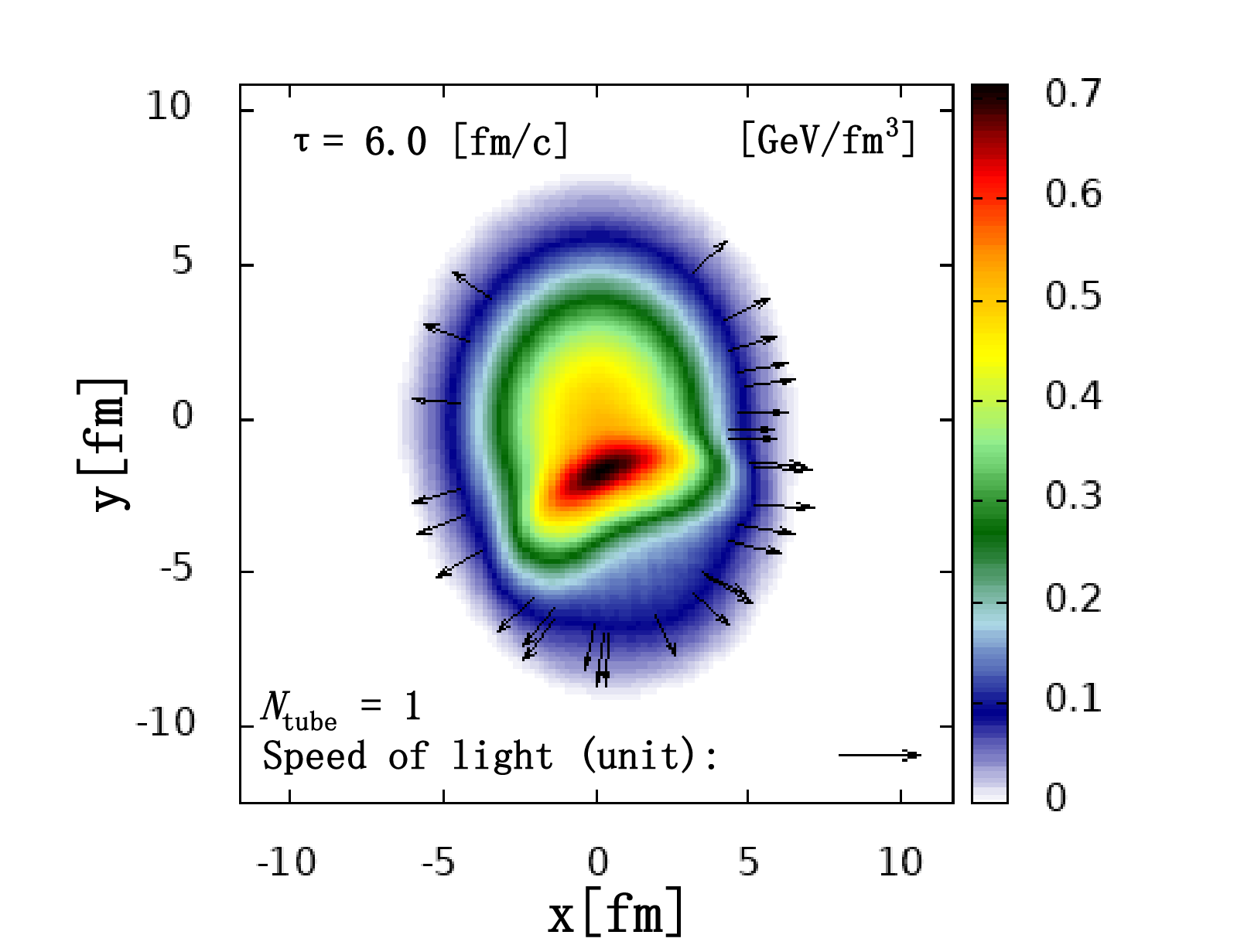}
    \end{minipage}
    \begin{minipage}{0.32\textwidth}
        \centering
        \includegraphics[width=1.15\textwidth, height=0.23\textheight]{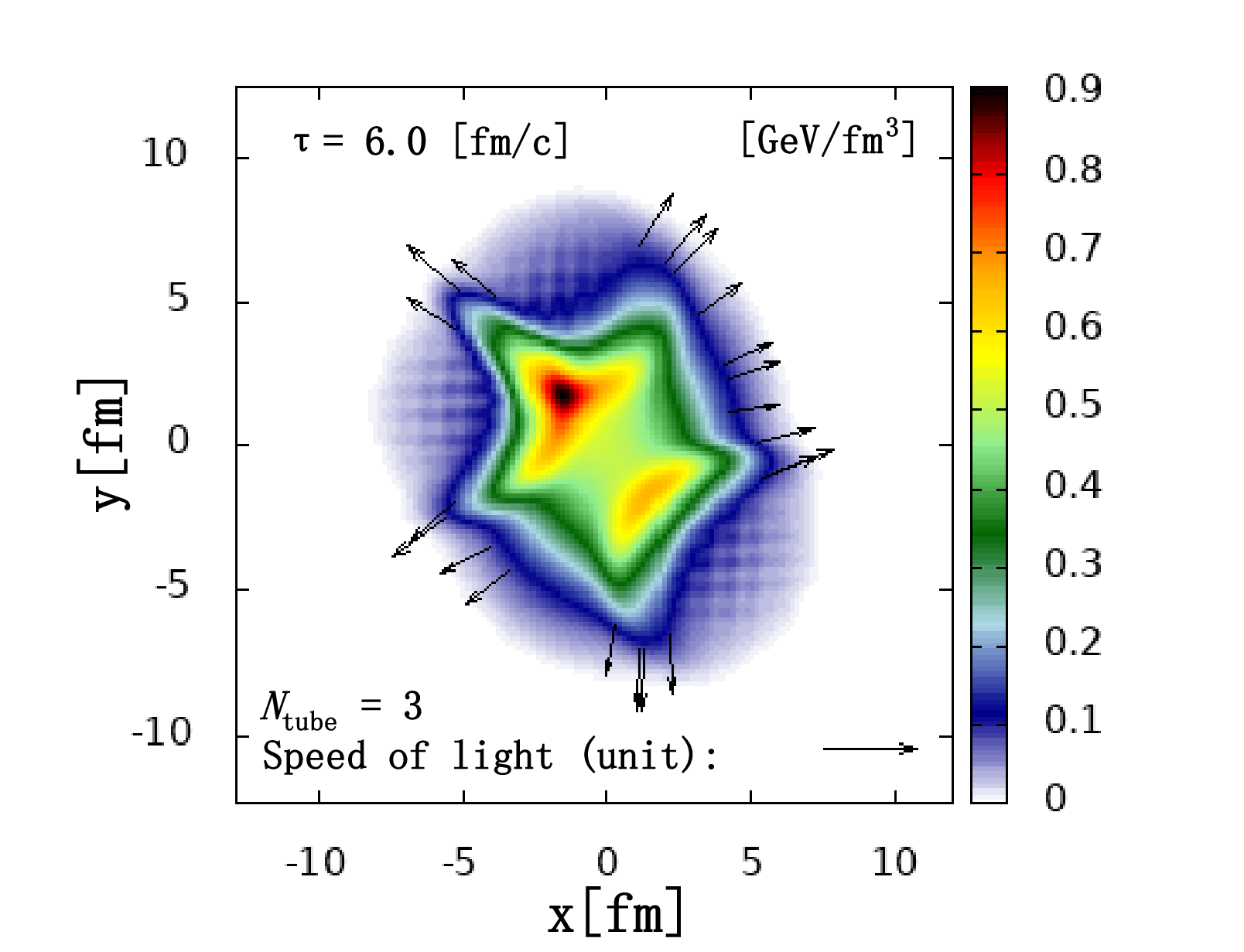}
    \end{minipage}
    \begin{minipage}{0.32\textwidth}
        \centering
        \includegraphics[width=1.15\textwidth, height=0.23\textheight]{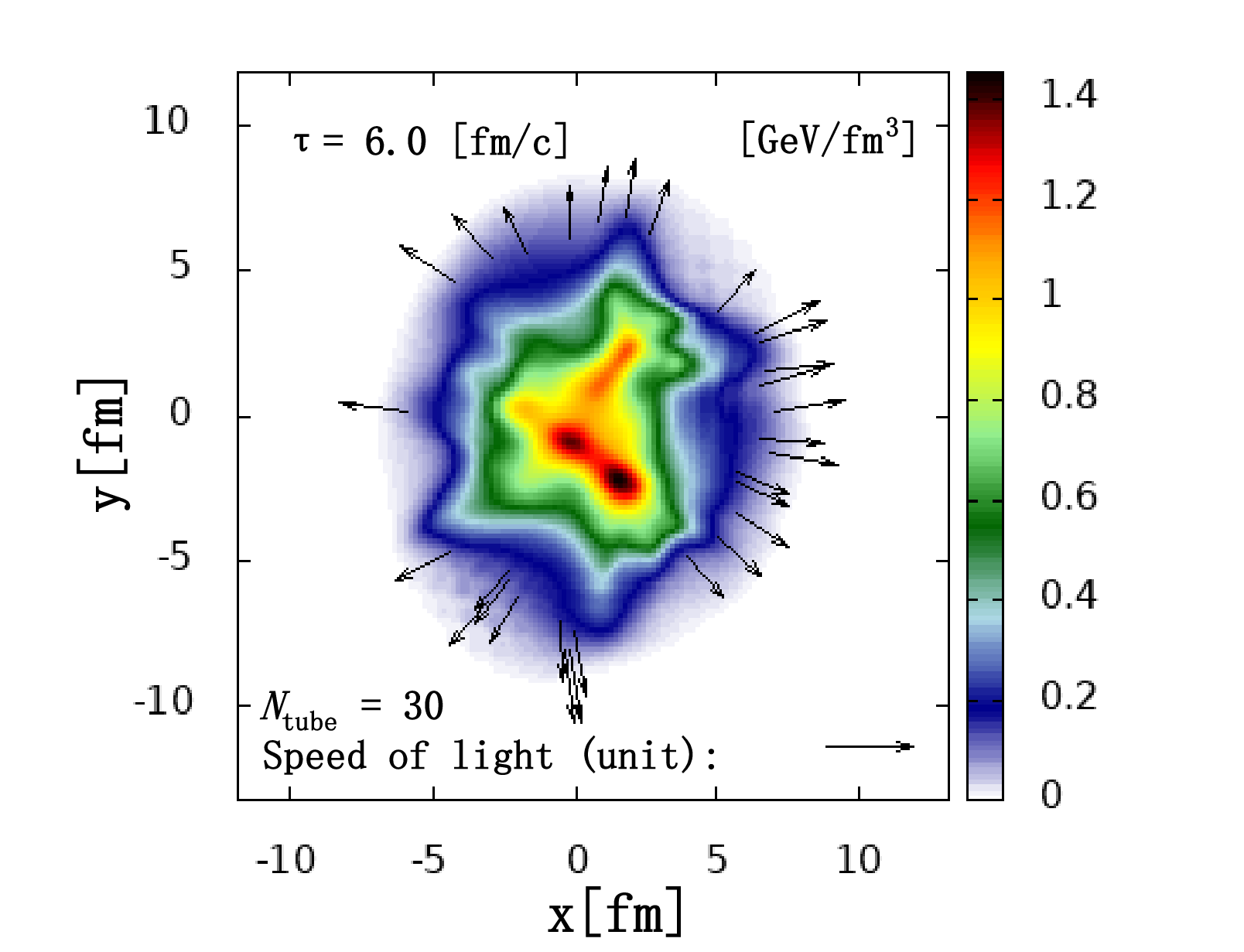}
    \end{minipage}
    \begin{minipage}{0.32\textwidth}
        \centering
        \includegraphics[width=1.15\textwidth, height=0.23\textheight]{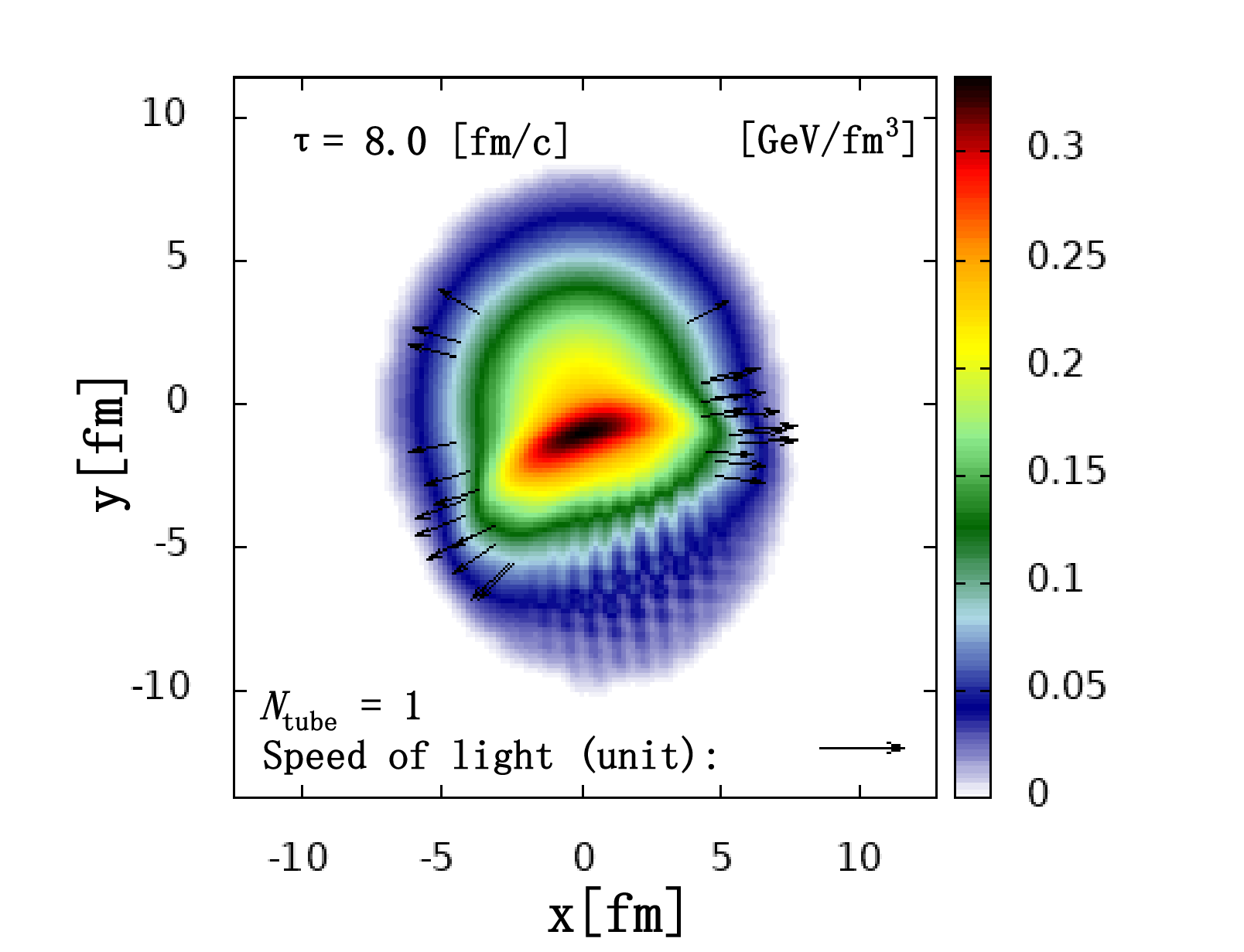}
    \end{minipage}
    \begin{minipage}{0.32\textwidth}
        \centering
        \includegraphics[width=1.15\textwidth, height=0.23\textheight]{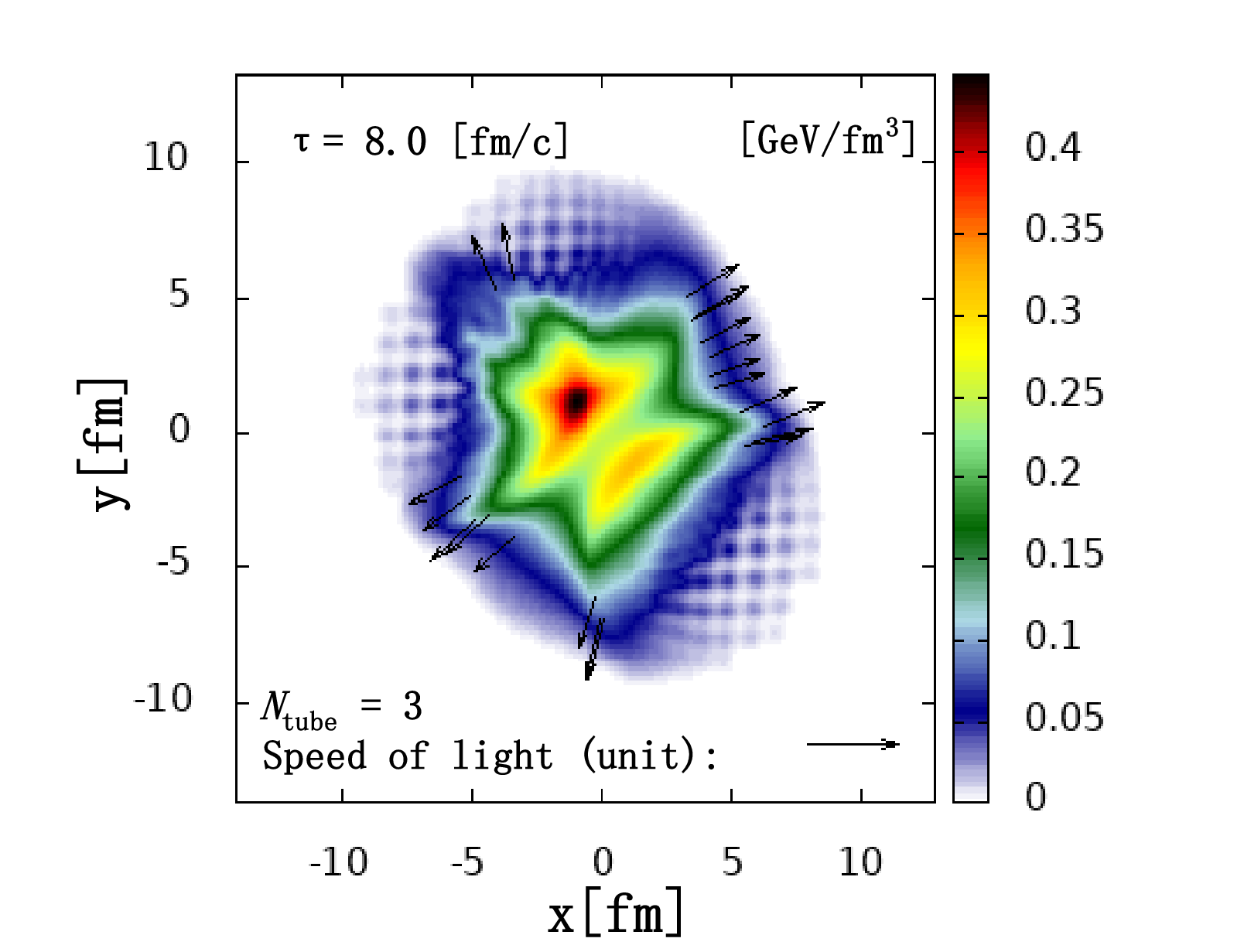}
    \end{minipage}
    \begin{minipage}{0.32\textwidth}
        \centering
        \includegraphics[width=1.15\textwidth, height=0.23\textheight]{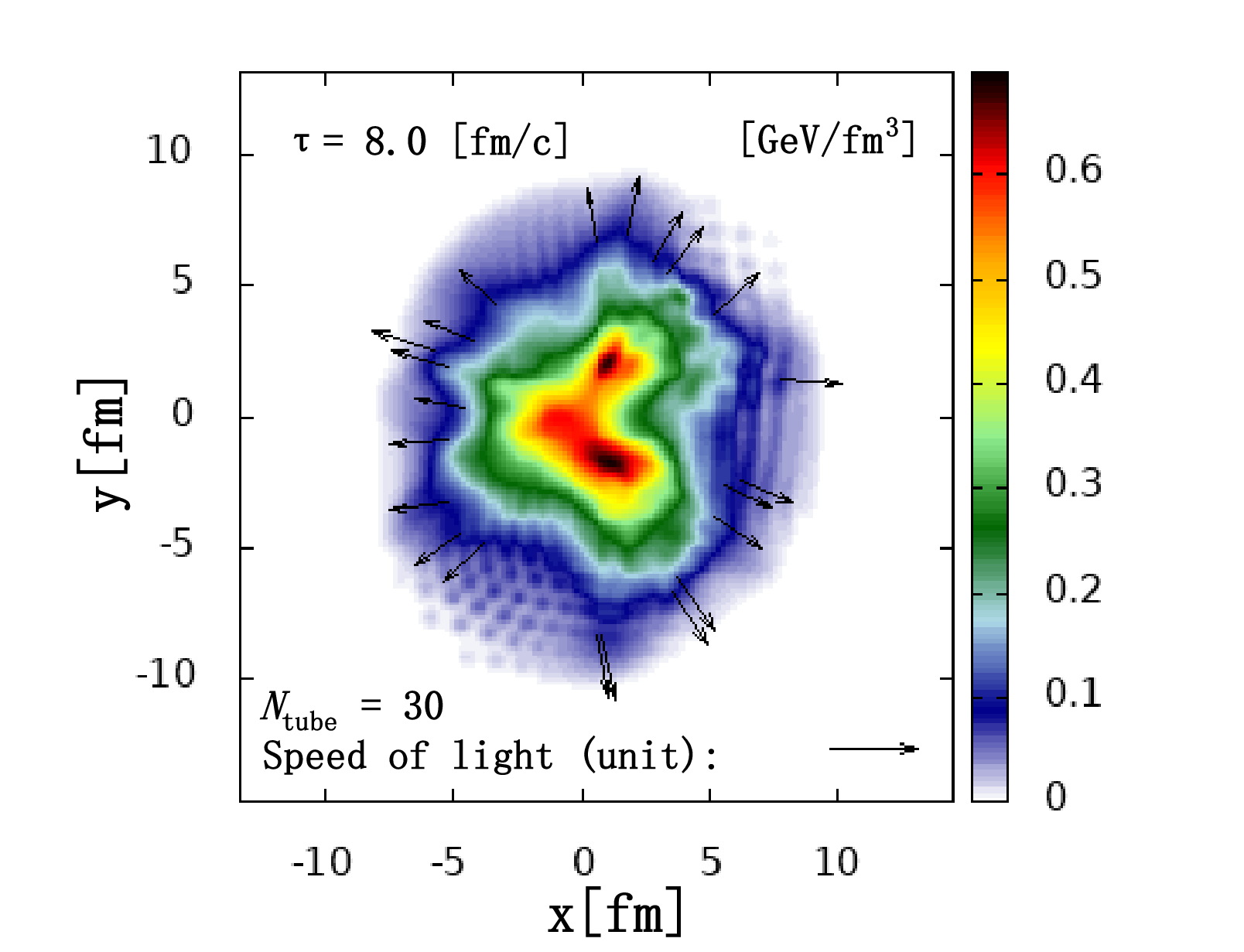}
    \end{minipage}
\caption{The temporal evolutions of different IC configurations with $N_\mathrm{tube} = 1$, $3$, and $30$ tubes.
The hydrodynamic simulations are carried out using the NeXSPheRIO code.
It is observed that the hydrodynamic expansion of the system can be drastically different for different IC.}
\label{fig_hydroevo}
\end{figure}

\section{Numerical results}\label{section4}

\begin{figure}[ht]
\centerline{\includegraphics[height=0.7\textwidth]{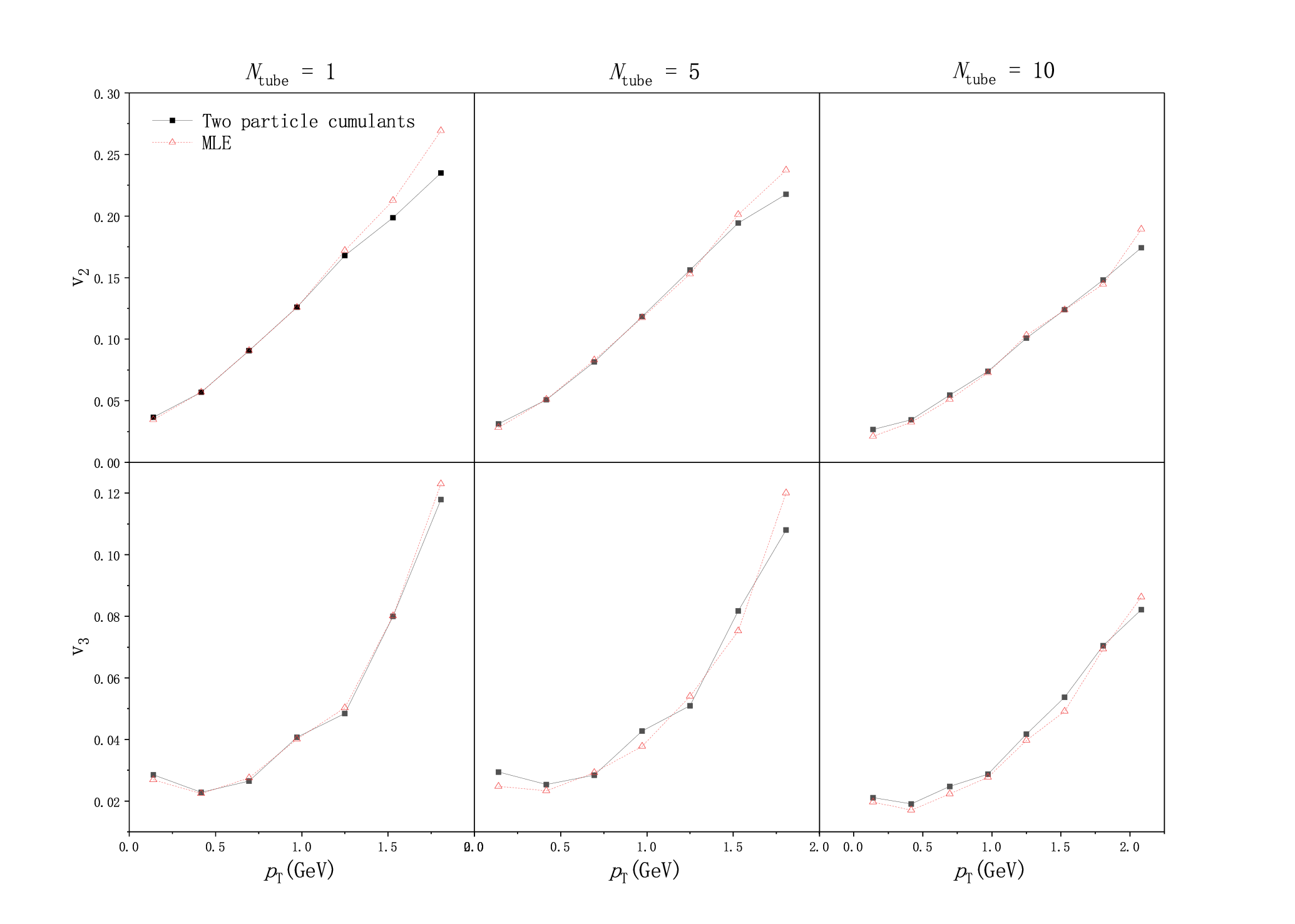}}
\renewcommand{\figurename}{Fig.}
\vspace{-0.5cm}
\caption{Event-by-event averaged elliptic and triangular differential flows for IC with different numbers of tube $N_\mathrm{tube}$.
The numerical calculations have been carried out using the MLE and particle cumulant methods.}
\label{fig_difflow}
\end{figure}

\begin{figure}[ht]
\centerline{\includegraphics[height=0.7\textwidth]{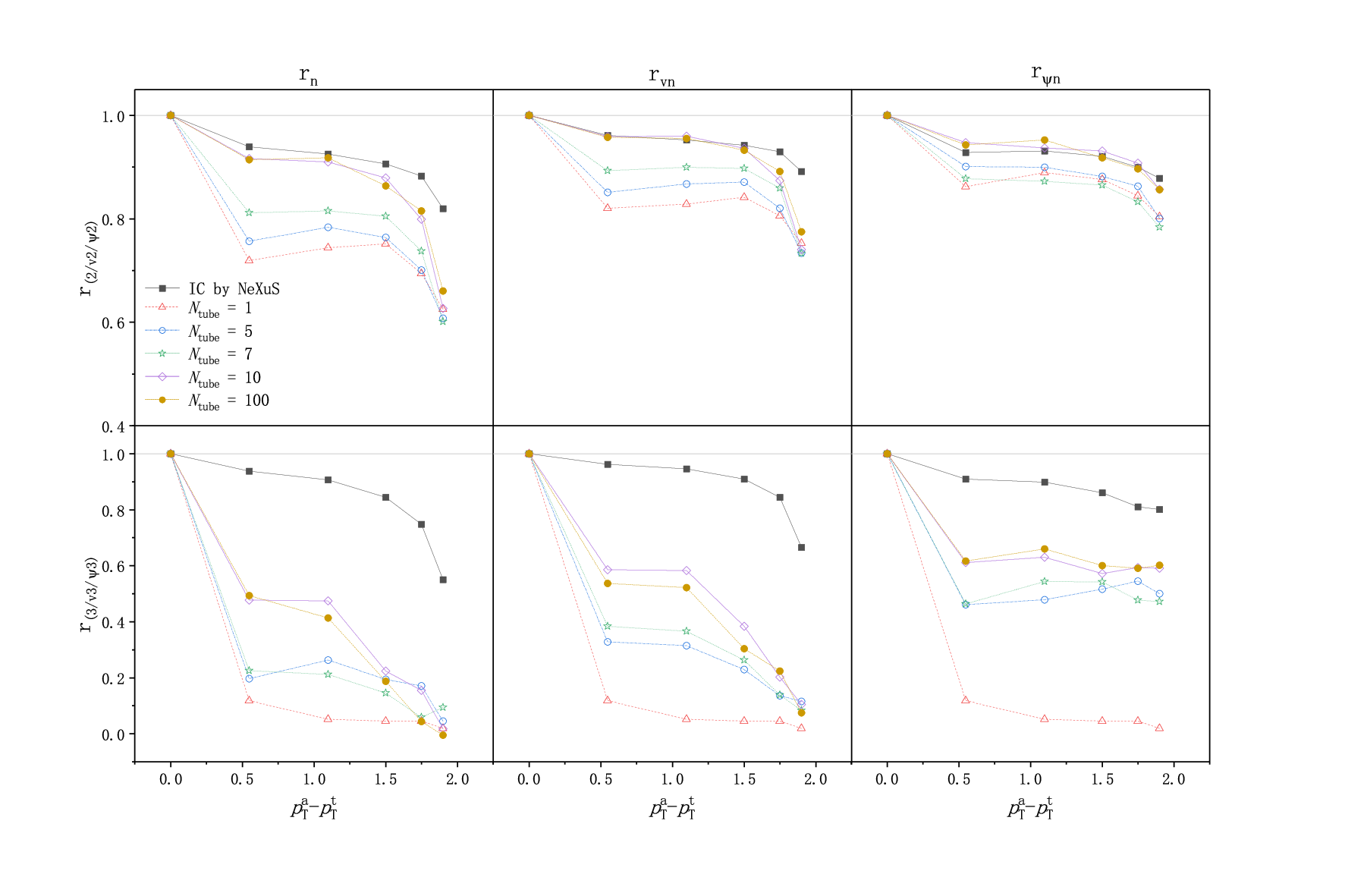}}
\renewcommand{\figurename}{Fig.}
\vspace{-0.5cm}
\caption{The obtained flow factorization ratio $r_n$, flow harmonics and event-plane correlations as functions of $p^{\rm a}_{\rm T}-p^{\rm t}_{\rm T}$ for event-by-event fluctuating IC generated by randomly casting $N_\mathrm{tube}=100$ peripheral tubes.
The results are also compared against event-by-event fluctuating IC generated by the NeXuS, shown by black filled squares.
The numerical calculations have been carried out using the MLE method.
The three columns correspond to factorization ratio, flow magnitude, and event-plane correlations from left to right.
The top row shows the results for $r_2$, while the bottom row displays the calculated $r_3$.}
\label{fig_factorization_vn}
\end{figure}

\begin{figure}[ht]
\centerline{
\includegraphics[height=0.4\textwidth]{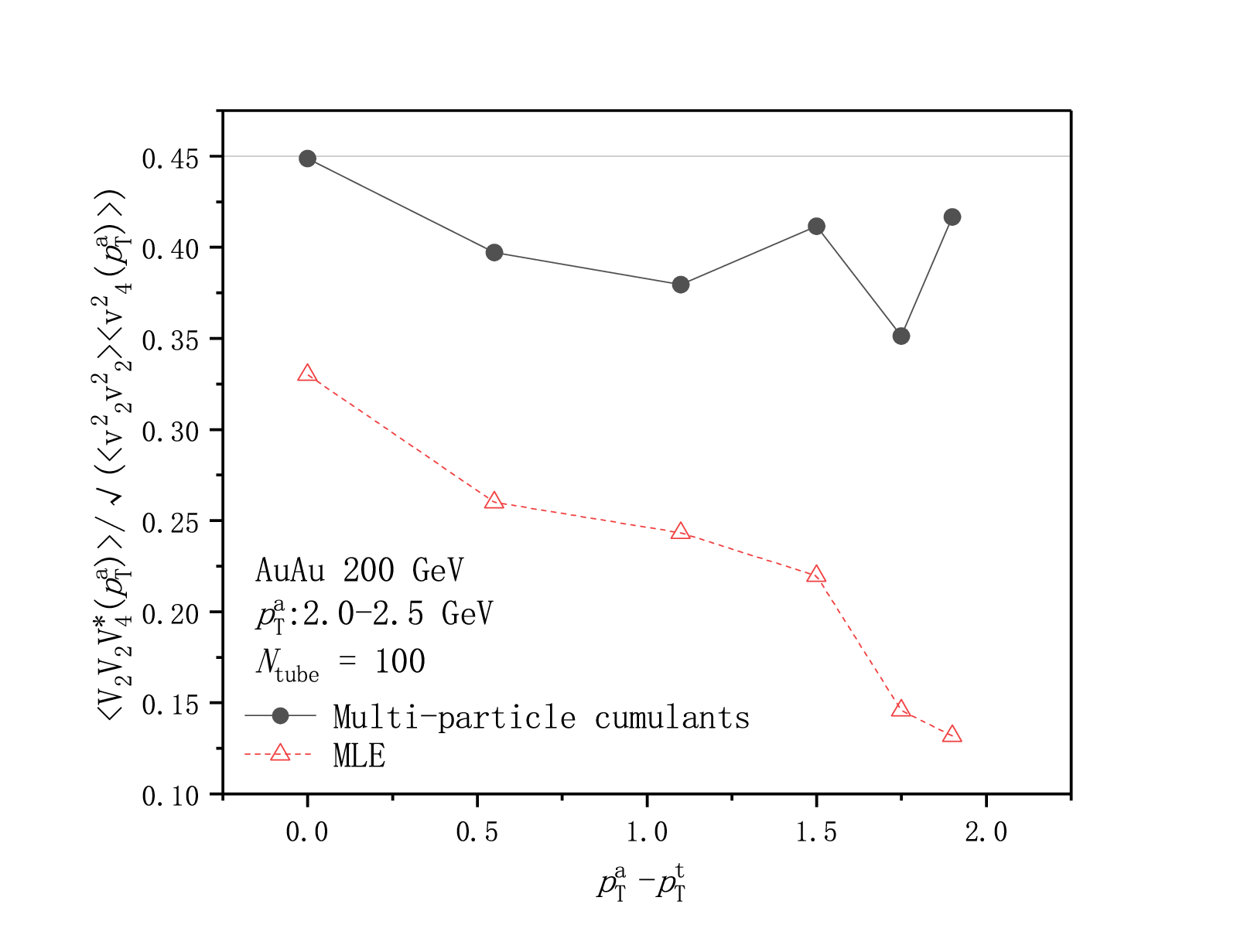}
\includegraphics[height=0.4\textwidth]{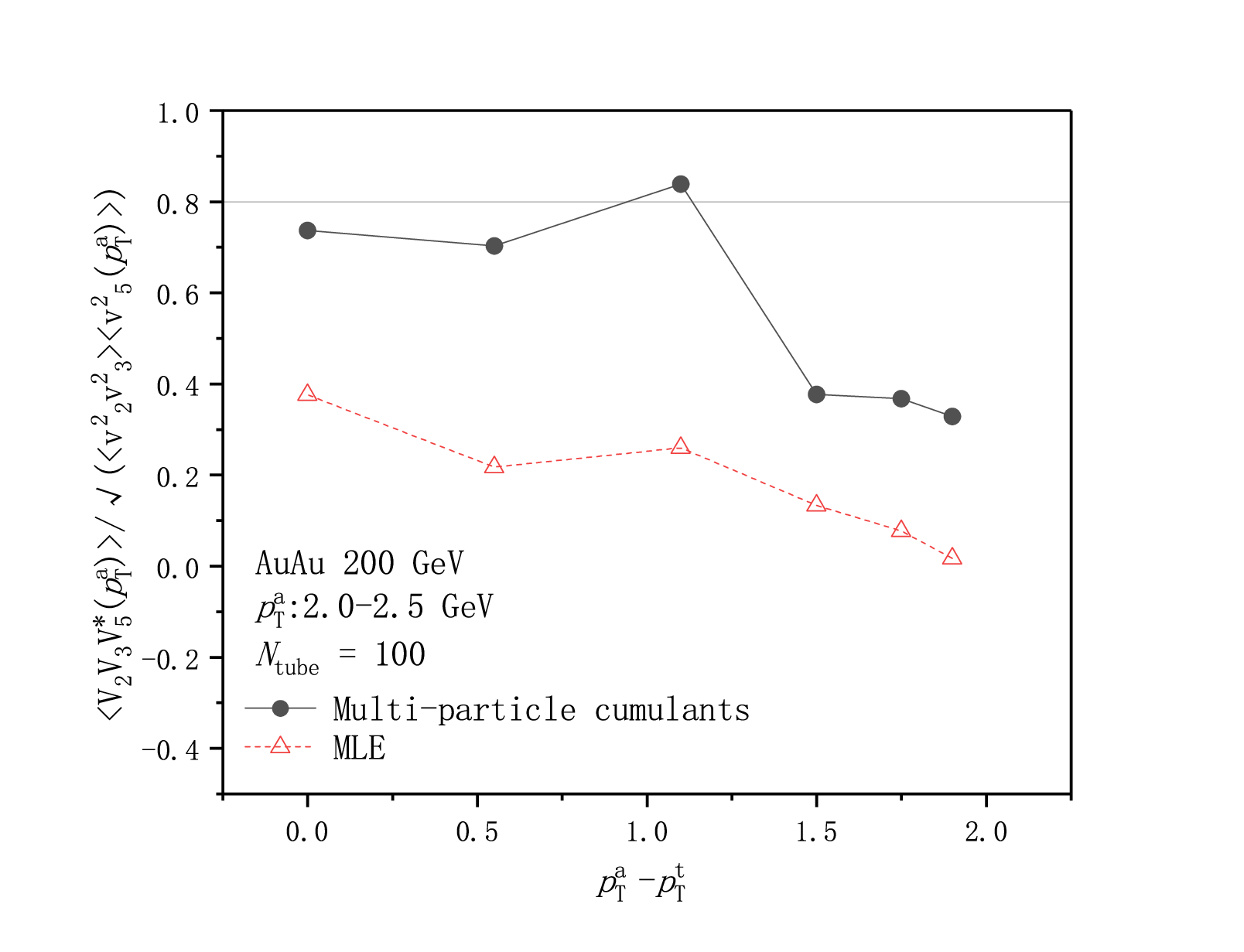}}
\renewcommand{\figurename}{Fig.}
\vspace{-0.5cm}
\caption{The mixed harmonic factorization ratios as functions of $p^{\rm a}_{\rm T}-p^{\rm t}_{\rm T}$ for event-by-event fluctuating IC generated by randomly casting $N_\mathrm{tube}=100$ peripheral tubes.
The left panel shows the ratio as a function of momentum interval for $m=5$, $k=3$, and $n=2$, and the right panel gives that for $m=4$, $k=2$, and $n=2$.
The calculations are carried out using the MLE and multi-particle cumulants.}
\label{fig_factorization_mvn}
\end{figure}

Using the devised IC furnished by the peripheral tube mode, the final-state particles are obtained by numerical simulations using the hydrodynamic code NeXSPheRIO~\cite{sph-review-01}.
We evaluate the temporal evolution, differential flow harmonics, and multi-particle correlations in terms of flow factorization.
The calculations are performed for IC configurations with different numbers of tube $N_\mathrm{tube}$.
The numerical results are shown in Figs.~\ref{fig_hydroevo}-\ref{fig_factorization_mvn}.

As shown in Fig.~\ref{fig_hydroevo}, the dynamic evolutions can drastically differ depending on different IC configurations.
As demonstrated in the left and middle columns of Fig.~\ref{fig_hydroevo}, when there are only a few tubes, the fluid is observed to be deflected in the vicinity of each tube to both sides.
In particular, the evolution around an individual high-energy tube leads to two peaks separated by roughly $120$ degrees in the azimuthal distribution, as indicated by the left-most column of Fig.~\ref{fig_hydroevo}.
Subsequently, this gives rise to the desired two-particle distribution where a double peak is formed on the away side, as pointed out in previous studies~\cite{sph-corr-12, sph-corr-18}.
However, as the number of tubes increases, the hydrodynamic evolution associated with different tubes becomes significantly interfered.
As shown in the right column, the resulting evolutions are rather complex.

However, the rather drastic difference in the initial conditions and hydrodynamic evolutions are not directly reflected in the flow harmonics, which measure the global anisotropies in the momentum space.
As indicated by Fig.~\ref{fig_difflow}, the resulting differential elliptic and triangular flow harmonics are mainly irrelevant to the difference in the IC.
Specifically, the resultant differential flows are rather robust against the variation of the number of tubes.
This result can be understood as the background primarily governs the overall elliptic and triangular shape of the IC, while the number of tubes impacts mainly the granularity.
Consistent with previous findings~\cite{sph-corr-20}, this result implies that the resulting two-particle correlation structure remains, by and large, unchanged.
Notably, the results shown in Fig.~\ref{fig_difflow} are carried out using two approaches, the MLE and the multi-particle cumulant.
On the one hand, the two-particle cumulant estimates flow harmonics through $v_n^2$ by Eq.~\eqref{eqEst2}.
By definition, the estimation is skewed owing to the presence of flow fluctuation~\cite{hydro-corr-ph-09, sph-corr-31}.
On the other hand, the MLE evaluates the flow harmonics by maximizing the likelihood, Eq.~\eqref{defMLE}, which also might lead to a biased estimation before attaining the statistical limit.
Nonetheless, it is observed that these two methods give consistent results as good agreements are manifestly reached for the differential flow.
Therefore, one must explore more sensitive quantities to scrutinize the observational impact from different granularities of the fluctuating IC.

In this regard, we proceed to evaluate flow factorization~\cite{ALICE:2011svq, Gardim:2012im}, which is more susceptible to initial state fluctuations.
For flow factorization, we elaborate on three different scenarios, and the results are presented in Figs.~\ref{fig_factorization_vn} and~\ref{fig_factorization_mvn}.
The numerical results indeed indicate that a significant difference is observed in such quantities. 
Specifically, we focus on two types of deviations.
First, we explore the flow factorization's dependence on the event's granularity by varying the number of tubes that constitute the IC. 
Secondly, a sizable difference is also observed across different flow estimators, namely, the MLE and multi-particle cumulants.
In other words, even though flow harmonics are found to be broadly consistent across different estimations, observables associated with the high-order moment of the one-particle distribution Eq.~\eqref{oneParDis} entail rather substantial deviations.

Specifically, we consider three types of factorization ratios.
The first scenario involves a ratio regarding two identical flow harmonics, as defined in Eq.~\eqref{rn_fact}.
This quantity is evaluated for elliptic and triangular flows by employing IC with different numbers of tube $N_\mathrm{tube}$.
As presented in the left column of Fig.~\ref{fig_factorization_vn}, the calculated factorization ratio $r_n$ is shown as a function of the difference between the transverse momenta of the trigger and associated particles.
The calculations are carried out using the MLE method.
At the origin, the two transverse momentum intervals of $p_{\rm T}^{\rm a}$ and $p_{\rm T}^{\rm t}$ coincide, and therefore, any substantial deviation from the unit comes solely from the correlation within the small given interval.
Our numerical calculations indicate that the deviation from perfect factorization mostly vanishes at the origin of the coordinates, consistent with the experimental data~\cite{CMS:2015xmx, CMS:2013bza}.
In theory, this is understood because at the limit, when the size of the interval vanishes, the Pearson correlation falls back to that between two identical quantities, which is guaranteed to have a perfect correlation.
Moreover, based on Eq.~\eqref{VnDeltaFactorization}, the flow vector can be further factorized if there were no fluctuations.
Even though such a factorization is not exact, it indicates that $r_n$ can be roughly viewed as receiving contributions from two factors: the Pearson correlation of flow harmonics and event-plane correlation.
These results are shown in the middle and right columns of Fig.~\ref{fig_factorization_vn}.
By comparing the three columns, it is observed that correlations between flow harmonics and event planes are both crucial in forming the resulting flow factorization.
The results obtained for different numbers of tubes indicated that a more significant breakdown of the factorization occurs when the number of tubes is small and gradually recovers when the number of tubes increases.
For a large number of tubes, the resulting IC is somewhat reminiscent of that of realistic IC generated by NeXuS~\cite{nexus-1, nexus-rept}.
This is particularly true for the elliptic flow, as the resulting factorization ratios approach that of Au+Au collisions, as indicated by black filled squares in Fig.~\ref{fig_factorization_vn}.
This result demonstrates that flow factorization is a more sensitive quantity to quantify the initial state fluctuations, precisely, the degree of granularity.

The second scenario involves a mix of three different flow harmonics of the form
\begin{eqnarray}
r_\mathrm{mix}(p_\mathrm{T}^a, p_\mathrm{T}^t, p_\mathrm{T}^t)&=&\frac{\langle V_m^*(p_\mathrm{T}^a)V_k(p_\mathrm{T}^t)V_n(p_\mathrm{T}^t)\rangle}{\sqrt{\langle V_m^*(p_\mathrm{T}^a) V_m(p_\mathrm{T}^a)\rangle \langle V_k^*(p_\mathrm{T}^t)V_k(p_\mathrm{T}^t) V_n^*(p_\mathrm{T}^t)V_n(p_\mathrm{T}^t)\rangle}} \nb\\
&=&\frac{\langle v_m(p_\mathrm{T}^a)v_k(p_\mathrm{T}^t)v_n(p_\mathrm{T}^t) \cos{(m \Psi_m(p_\mathrm{T}^a)-k \Psi_k(p_\mathrm{T}^t)-n \Psi_n(p_\mathrm{T}^t))} \rangle}{\sqrt{\langle v_m^2(p_\mathrm{T}^a)\rangle \langle v_k^2(p_\mathrm{T}^t) v_n^2(p_\mathrm{T}^t)\rangle}},
\label{rmix} 
\end{eqnarray}
where the three indices satisfy the relation
\begin{eqnarray}
m=k+n . 
\label{rmixRelation} 
\end{eqnarray}
Such a quantity has been investigated in the literature regarding flow fluctuations~\cite{Qian:2017ier, Bozek:2017thv}, as the term related to event-plane correlation readily vanishes if one has a global constant event plane. 
As the above quantity depends on three transverse momenta, in practice, one assigns two particles as triggers and one remaining particle as an associated particle, whose transverse momentum will be integrated over a given interval, as shown in the figures.
Here, we are also interested to compare the resulting factorization ratios using two different approaches: the MLE and particle cumulant methods.
The resulting flow factorization ratios are shown in Fig.~\ref{fig_factorization_mvn}, where one considers the range $2.0 < p_\mathrm{T}^a < 2.5$ GeV.

One observes that the factorization ratios obtained using two different methods possess a similar trend.
Nonetheless, the difference between these two methods is rather pronounced.
In addition to the analysis performed in Fig.~\ref{fig_factorization_vn}, this further reinforces our conclusion that factorization is a more sensitive quantity to the initial state fluctuations.
Besides that it distinguishes different degrees of granularity, the quantity is rather sensitive to the employed method, particularly when the quantity in question involves an explicit role of higher order moments.
In other words, the consistency between different approaches observed for differential flow in Eq.~\eqref{fig_difflow} does not generalize straightforwardly to factorization ratios.
Also, for such a scenario, it is noted that the factorization ratio does not traverse the origin of the coordinates.
This is understood as, in this case, the underlying correlation does not assume the Pearson correlation of the same quantity as its limit, and therefore the factorization ratio does not fall back to unity even when the momentum intervals coincide.

\begin{figure}[ht]
\centerline{
\includegraphics[height=0.4\textwidth]{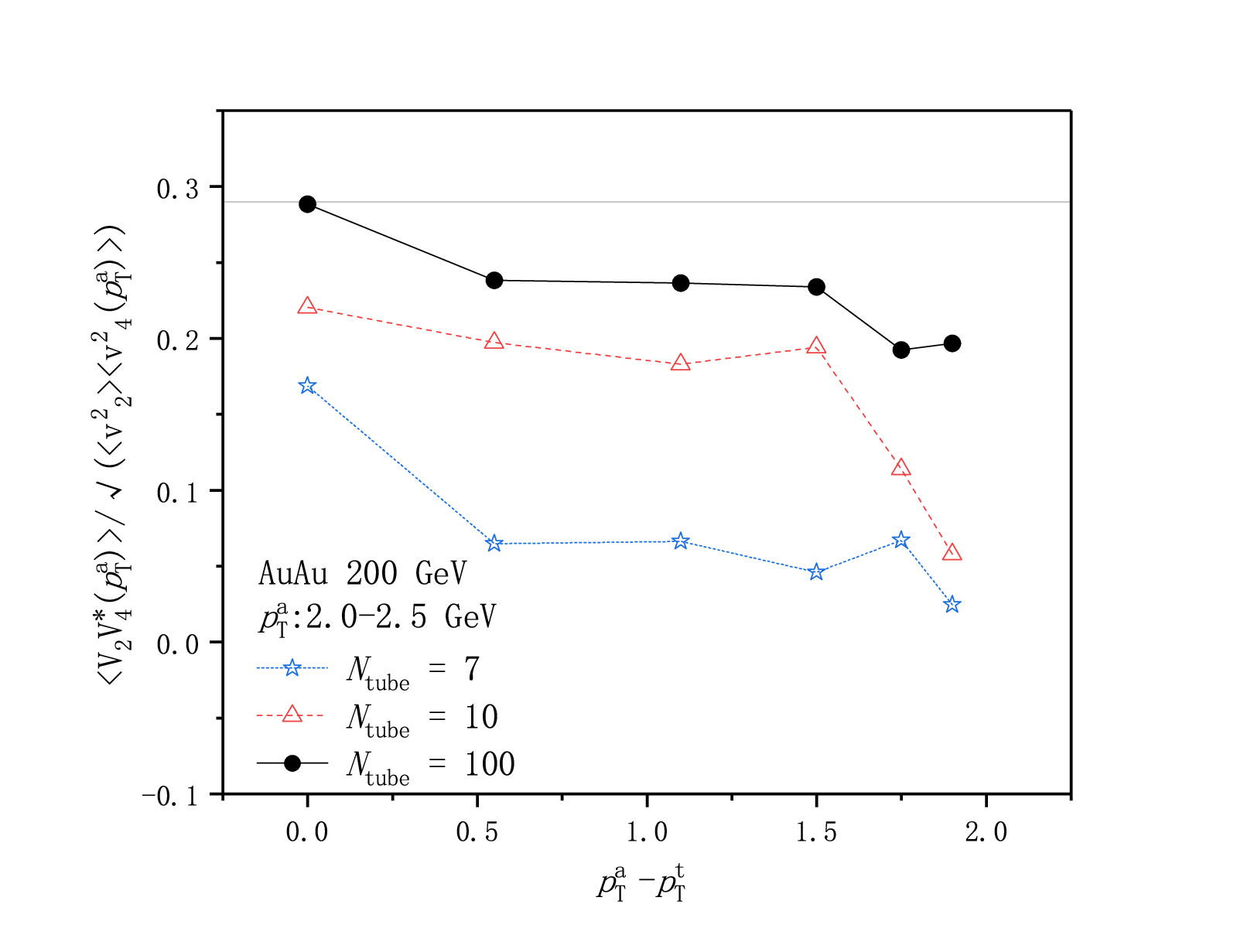}
\includegraphics[height=0.4\textwidth]{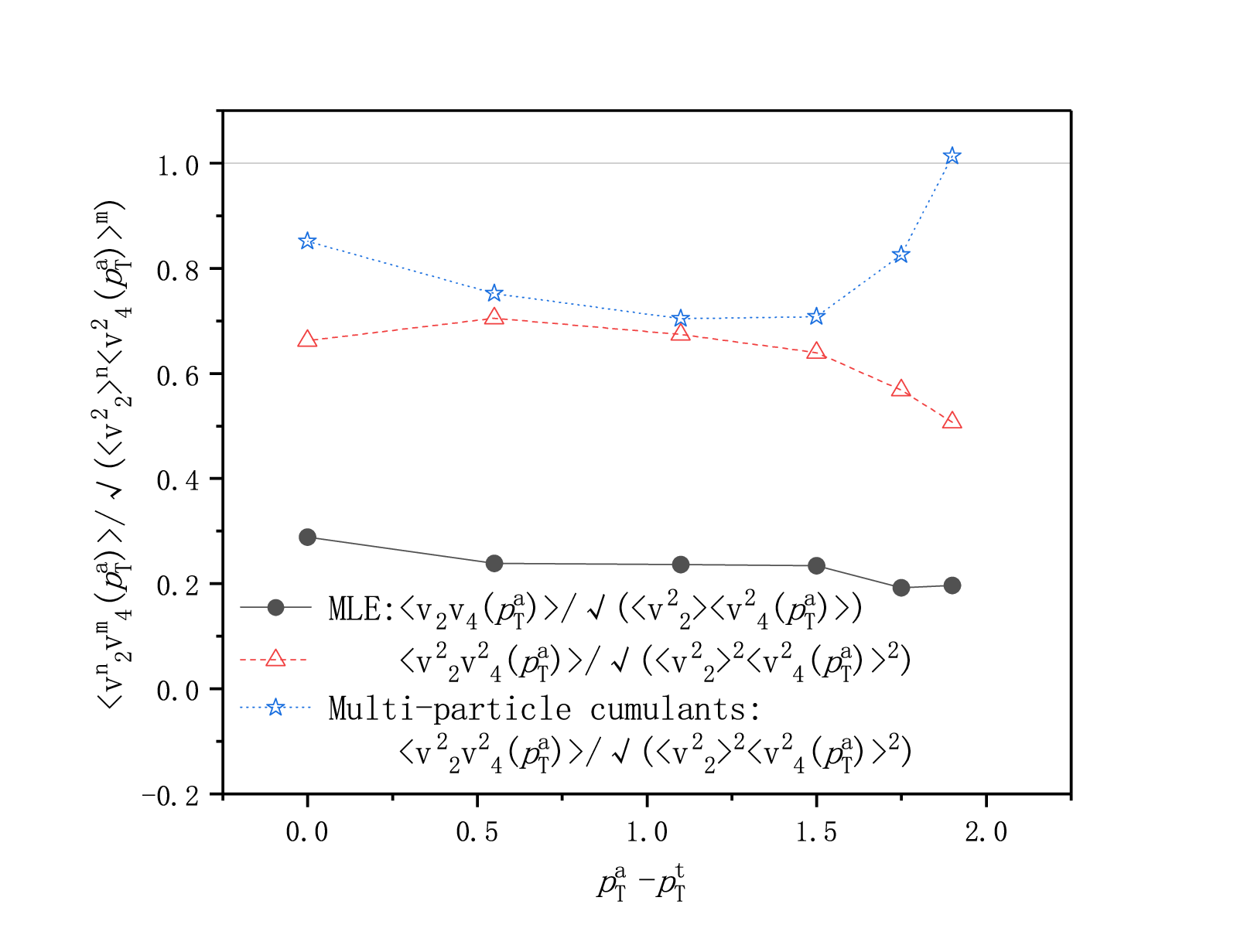}}
\renewcommand{\figurename}{Fig.}
\vspace{-0.5cm}
\caption{The mixed harmonic factorization ratios as functions of $p^{\rm a}_{\rm T}-p^{\rm t}_{\rm T}$ for the peripheral tube model, where the coefficients preceding the azimuthal angles do not satisfy the condition given by Eq.~\eqref{rmixRelation}.
We consider the specific choice of $m=4$ and $n=2$.
The left panel shows the results for IC generated using different numbers of tubes.
The right panel evaluates the same quantity for the IC associated with $N_\mathrm{tube}=100$ using various constructions in terms of different order of moments.
The calculations are carried out using the MLE and multi-particle cumulants.}
\label{fig_factorization_hvn}
\end{figure}

Lastly, we are interested in a further variation of the factorization ratio, essentially when the relation between indices Eq.~\eqref{rmixRelation} is no longer satisfied.
For this case, even if the event plane is a global constant, it will not be canceled out by the particle tuple combination in question.
As a result, such a case corresponds to a specific scenario where the MLE plays a more distinct role.
In the left panel of Fig.~\ref{fig_factorization_hvn}, we explore the dependence of the results on IC of different granularities furnished by different numbers of tubes.
A feature similar to Fig.~\ref{fig_factorization_vn} is observed, as the degree of factorization breaking decreases as the number of tubes increases.
Nonetheless, we note that it would still be possible to form particle pairs to assess the harmonics through the higher moments of the underlying one-particle distribution.
However, in such a way, we argue that the results might be quite sensitive to the specific construction.
This can be shown by explicitly evaluating their MLE counterparts in terms of high moments.
The calculations can be carried out straightforwardly owing to the {\it equivariance} of MLE~\cite{book-statistical-inference-Wasserman}.
The results are presented in the right panel of Fig.~\ref{fig_factorization_hvn}, for which the calculations are carried out for the IC generated using $N_\mathrm{tube}=100$ randomized tubes, corresponding to the black filled circles shown in the left panel.
For both panels, again, the factorization ratio does not traverse the origin of the coordinates.
It is observed that the use of different moments affects not only the magnitude of the factorization ratio but also its momentum dependence.
Even for the same construction, the results obtained by using the MLE and multi-particle cumulant methods are different.
In particular, as the MLE and particle correlators are not mathematically equivalent statistical estimators, such a difference is not unsurprising.
We, therefore, argue that the MLE provides a meaningful alternative to assess collective flows besides existing means.

\section{Concluding remarks}\label{section6}

Using the MLE, we investigated the relationship between initial-state fluctuations and final-state flow anisotropies in relativistic heavy-ion collisions. 
By reflecting on existing results that the mostly linear relation between initial state eccentricities and final state anisotropies~\cite{hydro-vn-34,hydro-vn-45}, the present study proceeds further based on the observation that distinct IC may produce almost identical momentum-space anisotropy in terms of flow harmonics~\cite{sph-corr-30}.
In this regard, our primary focus was on evaluating how the granularity of IC, modeled using a peripheral tube approach, impacts flow factorization, a more sensitive probe compared to traditional flow harmonics. 
Specifically, while differential flow harmonics, such as $v_2$ and $v_3$, showed minimal sensitivity to changes in the number and configuration of tubes, the flow factorization ratio displayed substantial variation, highlighting its potential as an effective observable for uncovering detailed information about the initial state.

By employing MLE, we extracted some specific mix-order correlators that are otherwise challenging to access, and the results showed distinct differences compared to standard techniques. 
The present study further generalizes our initial proposal~\cite{sph-vn-10, sph-vn-11} to a more specific subject.
A primary challenge of the approach was its computational cost.
Our findings indicate that such an extension is computationally feasible. 
Moreover, MLE provides a more nuanced understanding of flow factorization and its dependence on initial-state granularity.
As an asymptotically normal estimator, MLE's robustness and flexibility offer a new approach for analyzing complex multi-particle correlations. 
In particular, its implementation does not rely on constructing particle pairs or tuples to cancel event planes, substantially leading to broader applications.
Also, compared to the standard methods, it is more flexible to deal with scenarios where a template is not known prior to the analysis.
This feature indicates a promising aspect in flow analysis that MLE can be applied.
We plan to continue exploring related topics in future studies.

\section*{Acknowledgements}

We are thankful for enlightened discussions with Sandra Padula, Takeshi Kodama, and Yogiro Hama.
We gratefully acknowledge the financial support from Brazilian agencies 
Funda\c{c}\~ao de Amparo \`a Pesquisa do Estado de S\~ao Paulo (FAPESP), 
Funda\c{c}\~ao de Amparo \`a Pesquisa do Estado do Rio de Janeiro (FAPERJ), 
Conselho Nacional de Desenvolvimento Cient\'{\i}fico e Tecnol\'ogico (CNPq), 
and Coordena\c{c}\~ao de Aperfei\c{c}oamento de Pessoal de N\'ivel Superior (CAPES).
A part of this work was developed under the project Institutos Nacionais de Ciências e Tecnologia - Física Nuclear e Aplicações (INCT/FNA) Proc. No. 464898/2014-5.
This research is also supported by the Center for Scientific Computing (NCC/GridUNESP) of S\~ao Paulo State University (UNESP).
SFS acknowledges the funding from the Physics Master Teacher Studio of Guangsha University.
CY acknowledges the support of the Postgraduate Research \& Practice Innovation Program of Jiangsu Province under Grant No. KYCX22-3453.
JL acknowledges the support of the National Natural Science Foundation of China under Grant No. 12347101.

\bibliographystyle{h-physrev}
\bibliography{references_MLE, references_qian}

\end{document}